\documentclass[lettersize,journal]{IEEEtran}
\usepackage{amsmath,amsfonts}
\usepackage{algorithm}
\usepackage{array}
\usepackage[caption=false,font=normalsize,labelfont=sf,textfont=sf]{subfig}
\usepackage{textcomp}
\usepackage{stfloats}
\usepackage{url}
\usepackage{verbatim}
\usepackage{graphicx}
\usepackage{caption}
\usepackage{cite}
\usepackage{amssymb, mathtools, bm, bbm}
\usepackage{booktabs}
\usepackage[mathscr]{eucal}
\newcommand\tran{\mkern-0.5mu\raise1.25ex\hbox{$\scriptscriptstyle\mathsf{T}$}\mkern-3mu}
\newcommand\ctran{\mkern-0.5mu\raise1.25ex\hbox{$\scriptscriptstyle\dagger$}\mkern-3mu}
\newcommand\numberthis{\addtocounter{equation}{1}\tag{\theequation}}

\DeclareMathAlphabet{\mathdutchcal}{U}{dutchcal}{m}{n}
\SetMathAlphabet{\mathdutchcal}{bold}{U}{dutchcal}{b}{n}
\DeclareMathAlphabet{\mathdutchbcal}{U}{dutchcal}{b}{n}
\usepackage{algorithm}
\usepackage[rightComments=true,commentColor=blue, beginComment=/*~,endComment=/*~,beginLComment=/*~,endLComment=/*~,italicComments=false]{algpseudocodex}

\allowdisplaybreaks
\begin{document}

\title{Block Orthogonal Sparse Superposition Codes for $\sf{L}^3$ Communications: Low Error Rate, Low Latency, and Low Transmission Power}

\author{Donghwa~Han,
        Bowhyung~Lee,
        Min~Jang,~\IEEEmembership{Member,~IEEE,}
        Donghun~Lee,
        Seho~Myung,
        and~Namyoon~Lee,~\IEEEmembership{Senior~Member,~IEEE}
        \thanks{A preliminary version of this paper was presented in part at the 2022 IEEE International Conference on Communications (ICC) \cite{BOSS_MMSE_A_MAP, BOSS_MMSE_A_MAP_Demo}.}
        \thanks{D. Han, M. Jang, and D. Lee are with Samsung Research, Samsung Electronics, Seoul 06765, South Korea (e-mail: donghwa.han@samsung.com; mn.jang@samsung.com; d.hoonie.lee@samsung.com).}
        \thanks{B. Lee is with Network Business, Samsung Electronics, Suwon, Gyeonggi 16677, South Korea (e-mail: bowhyung.lee@samsung.com).}
        \thanks{S. Myung is with Mobile eXperience (MX) Business, Samsung Electronics, Suwon, Gyeonggi 16677, South Korea (e-mail: seho.myung@samsung.com).}
        \thanks{N. Lee is with the School of Electrical Engineering, Korea University, Seoul 02841, South Korea (e-mail: namyoon@korea.ac.kr).}
}




\maketitle

\begin{abstract}

Block orthogonal sparse superposition (BOSS) code is a class of joint coded modulation methods, which can closely achieve the finite-blocklength capacity with a low-complexity decoder at a few coding rates under Gaussian channels. However, for fading channels, the code performance degrades considerably because coded symbols experience different channel fading effects. In this paper, we put forth novel joint demodulation and decoding methods for BOSS codes under fading channels. For a fast fading channel, we present a minimum mean square error approximate maximum a posteriori (MMSE-A-MAP) algorithm for the joint demodulation and decoding when channel state information is available at the receiver (CSIR). We also propose a joint demodulation and decoding method without using CSIR for a block fading channel scenario. We refer to this as the non-coherent sphere decoding (NSD) algorithm. Simulation results demonstrate that BOSS codes with MMSE-A-MAP decoding outperform CRC-aided polar codes, while NSD decoding achieves comparable performance to quasi-maximum likelihood decoding with significantly reduced complexity. Both decoding algorithms are suitable for parallelization, satisfying low-latency constraints. Additionally, real-time simulations on a software-defined radio testbed validate the feasibility of using BOSS codes for low-power transmission.\end{abstract}


\section{Introduction}
\subsection{Channel Coding Requirements for Next-Generation Wireless Networks}
\IEEEPARstart{U}{ltra} reliable, low-latency communication (URLLC) has extended its applicability across diverse domains, since its inception by IMT-2020 as a cornerstone of 5G NR solutions \cite{URLLC_Introduction_1, URLLC_Channel_Coding_2}. These applications include, but are not confined to, autonomous vehicles, traffic management, and smart power grids \cite{URLLC_Application_1_Tactile, URLLC_Application_2_Autonomous_Vehicle}. Moreover, network operators have witnessed a great surge in the demand for ultra-reliable, low-latency connectivity, driven by the proliferation of mission-critical applications and the emergence of new use cases \cite{URLLC_Application_3_Misstion_Critical}. Time-sensitive operations such as full factory automation, remote surgery through robotic interactions, and interactive holographic communications all translate to the necessity for ultra-high accuracy and exceedingly short latencies \cite{URLLC_Application_4_Holographic ,URLLC_Application_5_Healthcare}.



In response to this evolving market landscape, IMT-2030 has introduced hyper reliable and low-latency communication (HRLLC) as an extension of URLLC, with even more stringent requirements in terms of reliability (on the order of $10^{-7}$) and latency (ranging from 0.1 to 1 ms) \cite{IMT-2030}. Given the pivotal role of the physical layer (PHY) in meeting these exacting demands, channel coding techniques must evolve accordingly \cite{URLLC_Channel_Coding_1}. Low-rate transmission will stand as a linchpin in ensuring reliability within the HRLLC environment. By allocating increased resources and redundancy per information bit, the resilience of the the communication link against noise and fading impairments can be fortified. Additionally, the inherent latency constraints necessitate encapsulating information within short packets to minimize transmission overhead.

Concurrently, alongside the expanding deployment of 5G NR networks, the features introduced in 5G Advanced with Rel-18 are poised to mature and advance further in Rel-19 in tandem. A notable update is the inclusion of Ambient Internet of things (IoT) as a study item, providing impetus to standardization efforts \cite{AmbientIoT_approval}. Industry voices have emphasized the imperative for a new PHY design, including channel coding tailored for extremely low-power devices: either battery-less or with limited power storage \cite{AmbientIoT_Nokia, AmbientIoT_OPPO, AmbientIoT_Spreadtrum, AmbientIoT_ZTE}. A simple encoding scheme is essential to meet the requirements of low transmission power and low complexity. In this context, the compact nature of short packets dovetails with the tight energy constraints inherent to IoT devices. Furthermore, complex channel estimation based on resource-consuming pilot symbols may contribute to significant overhead. Consequently, detection schemes without the reliance on channel state information (CSI), i.e., non-coherent detection, are likely to be embraced for decoding channel codes in Ambient IoT.

As the transmission of low-rate, short data packets is expected to become the norm in the realm of HRLLC and Ambient IoT, an innovative coding scheme capable of delivering exceptional reliability and minimal latency, while operating within stringent power constrains, becomes imperative \cite{URLLC_Channel_Coding_2}.

\subsection{Block Orthogonal Sparse Superposition Codes}
Sparse superposition codes, also known as sparse regression codes (SPARCs), were introduced by Joseph and Barron for reliable communication over the additive white Gaussian noise (AWGN) channel \cite{Barron12_SPARC_adaptive_successive_decoding}. A SPARC is defined by a Gaussian dictionary matrix $A$, where codewords are linear combinations of its columns corresponding to the few non-zero elements of $\beta$. That is, a codeword can be expressed as $A \beta$, and transmitted information is encoded into the locations of the non-zero entries in $\beta$. SPARCs bridge error correction coding and compressed sensing (CS) \cite{Candes06_CS}, two seemingly unrelated domains, in that decoding SPARCs is equivalent to estimating $\beta$ from a noisy linear measurement of $A \beta$ by leveraging prior knowledge of the dictionary matrix $A$ (or sensing matrix in the language of CS). While prior research has proven that SPARCs can asymptotically achieve the AWGN capacity with various decoders \cite{Barron14_SPARC_ML_decoding, Barbier17_SPARC_AMP, Rush17_SPARC_AMP, Barbier19_SC_SPARC}, recent efforts have focused on improving empirical finite-length performance of SPARCs. This includes strategies such as wise power-allocation techniques \cite{Greig2018_SPARC_Finite}, list decoding \cite{Hao21_SPARC_List_Decoding}, and careful dictionary design \cite{Hsieh21_SPARC_FFT}. However, for short blocklength under a few hundred, a primary focus of this paper, the performance of SPARCs remains sub-optimal. This limitation hinders their potential application in real-word wireless communication scenarios. We address this limitation by investigating a special class of SPARCs in fading scenarios.

The authors have introduced block orthogonal sparse superposition (BOSS) codes as a novel class of SPARCs \cite{BOSS_AWGN_conf, BOSS_AWGN_journal}. An important distinction from SPARCs is that a BOSS code utilizes a structured dictionary matrix constructed as a concatenation of different unitary matrices. Information bits are encoded into the selection of a sub-dictionary matrix and its non-overlapping columns, expressing the codeword as a weighted sum of orthogonal columns of the chosen unitary matrix. The proposed two-stage maximum a posteriori (MAP) decoder leverages the mutual orthogonality of columns that comprise the codeword. Equipped with two-stage MAP decoding, BOSS codes were shown to outperform state-of-the-art cyclic redundancy check (CRC)-aided polar (CA-polar) \cite{Arikan09_Polar} and polarization-adjusted convolutional (PAC) codes \cite{Ariakn19_PAC} in the low-rate, short blocklength regime. Notably, when concatenated with the CRC outer code, BOSS codes achieve performance within one dB of the finite-length information-theoretic lower bound \cite{Polyanskiy10_dispersion_bound}. Moreover, \cite{Lee24_BOSS_ISSCC} showcased an integrated BOSS decoder fabricated in 28nm CMOS technology. This prototype not only boasts over 6 times lower latency compared to a polar counterpart but also occupies a minimal core area and demonstrates exceptional energy efficiency. These encouraging outcomes attest to the potential of BOSS codes in enabling low-latency and energy-efficient communications.

\subsection{Contributions}
The major contributions of this paper are summarized as follows:
\begin{itemize}
    \item  We first consider a transmission of BOSS codes in a single-input single-output (SISO) fast fading channel, in which each coded symbol experiences different fading coefficients such as the orthogonal frequency division multiplexing (OFDM) transmission systems. In this case, we put forth a joint demodulation and decoding method employing channel state information (CSI) at receiver. We refer to this as the minimum mean square error approximate-MAP (MMSE-A-MAP). Treating each coded symbol as an independent Gaussian prior, we derive an MMSE equalizer customized to the transmitted codeword rather than the original message vector that needs to be recovered. The proposed MMSE-A-MAP decoder lends itself well to parallelization, benefiting from the equalizer’s independence from the dictionary matrix. With the MMSE-A-MAP decoding approach, BOSS codes significantly outperform 5G CA-polar codes under the fast fading channels.
    \item   We also consider an uplink single-input multiple-output (SIMO) channel, where transmitted codeword symbols experience identical channel fading states, i.e., a block fading scenario. In this scenario, we introduce a non-coherent decoding method known as non-coherent sphere decoding (NSD) for pilot-free communications. NSD employs a two-stage quasi-maximum likelihood (ML) decoding architecture to enable parallelization. By treating the decorrelated output of each receiver antenna as jointly Gaussian with distinct variances, we derive a closed-form expression for the conditional joint density of these decorrelated outputs. This derivation facilitates a feasible quasi-ML decoding solution. The solution simplifies the decoding task to searching for the message vector most correlated with the decorrelated outputs. It allows for easy integration into the two-stage parallel ML framework, thereby minimizing latency.
    \item   Recognizing that the quasi-ML solution involves identifying the largest elements of the empirical average of the outer product of the decorrelated antenna output with itself, we propose an efficient sphere decoding algorithm for BOSS codes. This non-coherent sphere decoder generates a reduced search space of sub-message vector candidates by exploiting the structural characteristics of each outer product. Despite its reduced search space, the sphere decoder achieves performance comparable to that of the quasi-ML decoder, even with a small sphere size.
    \item   To verify the feasibility of the proposed MMSE-A-MAP algorithm, we implement a real-time software-defined radio (SDR) testbed. Experimental results demonstrate the possibility of reliable communications using BOSS codes with extremely low transmission power.
\end{itemize}

\subsection{Organization}
The remainder of this paper is organized as follows. Section II introduces two fading system models under consideration, followed by a brief introduction to encoding and two-stage MAP decoding of BOSS codes. Section III first addresses the limitations of two-stage MAP decoding in the SISO OFDM environment and proposes the MMSE-A-MAP algorithm. Section IV focuses on the uplink SIMO system and presents the two-stage ML decoding framework. Section V introduces the quasi-ML decoding algorithm and proposes the non-coherent sphere decoding method. Section VI provides numerical results of different decoding algorithms. Section VII discusses the SDR testbed and presents real-time simulation results. Finally, Section VIII concludes the paper by offering potential research directions.

\section{Preliminaries}
\subsection{System Model}
\textbf{AWGN channel:} The BOSS code was originally proposed for efficient, reliable communications over the AWGN channel. Upon sending a codeword $\mathbf{c} \in \mathbb{C}^M$, with $M$ being the blocklength, the complex received vector can be expressed as 
\begin{equation}
    \mathbf{y} = \mathbf{c} + \mathbf{v}, \label{AWGN channel model}
\end{equation} where $\mathbf{v}$ is the $M$-dimensional zero-mean circularly-symmetric multi-variate Gaussian vector with the common variance $\sigma_v^2$. That is, $\mathbf{v}$ has the covariance matrix $\sigma_v^2 \mathbf{I}_M$.

\textbf{SISO fast fading channel:} We study a Rayleigh multi-path fading channel with OFDM transmission. A diagonal matrix $\mathbf{\Lambda} = \text{diag} ( \lambda_1 , \lambda_2, \dots , \lambda_M ) \in \mathbb{C}^M$ represents the frequency-domain channel. Under the assumption that the length of cyclic prefix suffices to completely nullify inter-symbol interference (ISI), diagonal entries, $\lambda_i$, can be interpreted as ISI-free parallel channels, through each of which a modulated symbol propagates. With respect to a codeword $\mathbf{c}$, the observation at the receiver is given by
\begin{equation}
    \mathbf{y} = \mathbf{\Lambda} \mathbf{c} + \mathbf{v} , \label{SISO channel model}
\end{equation} where $\mathbf{v} \sim \mathcal{CN} ( \mathbf{0}_M, \sigma_v^2 \mathbf{I}_M )$.

\textbf{SIMO block fading channel:} We further investigate a BOSS-coded uplink system where a single-antenna sensor or small device transmits data to a receiver equipped with $N_\text{Rx}$ antennas. The complex channel vector $\bm{\eta} \in \mathbb{C}^{N_\text{Rx}}$ is defined as follows:
\begin{equation}
    \bm{\eta} = \begin{bmatrix}
        \eta_1 \\
        \eta_2 \\
        \vdots \\
        \eta_{N_\text{Rx}}
    \end{bmatrix} = \sqrt{\zeta} \mathbf{h} = \sqrt{\zeta} \begin{bmatrix}
        h_1 \\
        h_2 \\
        \vdots \\
        h_{N_\text{Rx}}
    \end{bmatrix} , \label{uplink channel model}
\end{equation} where $\zeta$ is the large-scale fading factor, and $h_n \sim \mathcal{CN}(0, 1)$. In essence, the transmitted signal arriving at each antenna propagates through an instantaneous realization of flat fading. 

Upon transmission of the codeword $\mathbf{c}$, the $n$-th antenna's received vector, $\mathbf{y}^{[n]} \in \mathbb{C}^M$, is given by
\begin{equation}
    \mathbf{y}^{[n]} = \eta_n \mathbf{c} + \mathbf{v}_n = \sqrt{\zeta} h_n \mathbf{c} + \mathbf{v}_n, \label{each antenna received vector}
\end{equation} where $\mathbf{v}_n \sim \mathcal{CN}(\mathbf{0}_M, \sigma_v^2 \mathbf{I}_M )$ is the noise vector at the $n$-th antenna. Entries of $\mathbf{v}_n$ are assumed to be independent of $\eta_n$.

While we acknowledge the use of a single symbol $\mathbf{y}$ for the received vector across different scenarios as an abuse of notation, distinctions can be clarified within the respective contexts.

\subsection{Successive BOSS Encoding}
Unlike conventional channel codes, the BOSS code is a joint coded-modulation scheme that encodes information bits directly into coded symbols for transmission without posterior modulation mapping. Let $G \in \mathbb{Z}^+$ be the block number. A BOSS code is defined by a dictionary matrix $\mathbf{A} \in \mathbb{C}^{M \times N}$, which is a concatenation of $G$ $M \times M$ unitary matrices:
\begin{equation}
    \mathbf{A} = \begin{bmatrix}
        \mathbf{U}_1 & \mathbf{U}_2 & \cdots & \mathbf{U}_G
    \end{bmatrix} .
\end{equation} It is obvious that $N = G \cdot M$.

Let $\mathbf{x} \in \mathbb{C}^N$ be a spare message vector with $K$ non-zero entries, i.e., $\| \mathbf{x} \|_0 := \text{supp}(\mathbf{x}) = K(\ll N)$. We further impose a block-sparsity constraint on $\mathbf{x}$ such that all non-zero coefficients are located in a single length-$M$ segment of $\mathbf{x}$. For example, assuming the $g$-th segment has meaningful elements, we have $\| \mathbf{x} \|_0 = \| \mathbf{x}_g \|_0 = K$ with $\mathbf{x}_g = [ x_{(g-1)\cdot M + 1}, x_{(g-1) \cdot M + 2} , \dots , x_{g \cdot M} ]$. 

The information bit-sequence is partitioned into separate blocks which are sequentially mapped into the message vector $\mathbf{x}$. First, the encoder uses $B_0 = \lfloor \log_2 (G) \rfloor$ bits to select a sub-dictionary matrix index $g$. Then, at each layer, the encoder uses $B_{\ell, \text{loc}}$ bits to choose $K_\ell$ columns of $\mathbf{U}_g$ from a candidate set $\mathscr{M}^{(\ell)}$. The next $B_{\ell, \text{val}}$ bits are used to select the ordered combination of $K_\ell$ values from the $\ell$-th layer's alphabet, $\mathcal{A}_\ell = \{ \alpha_{\ell, 1}, \alpha_{\ell, 2} , \dots, \alpha_{\ell, K_\ell} \}$, and these symbols are assigned to the positions of $\mathbf{x}_g$, corresponding to the selected column indices. We introduce a new notation, $\mathbf{x}_g^{(\ell)} \in \mathbb{C}^M$, for a constituent vector of $\mathbf{x}_g$ such that it contains only $K_\ell$ non-zero elements assigned at the $\ell$-th layer. The values of $K_\ell$ satisfy the sum constraint such that $\sum_{\ell=1}^L K_\ell = K$. We further define the non-zero support of $\mathbf{x}_g^{(\ell)}$ as $\mathcal{I}^{(\ell)} := \{ m \in \mathscr{M}^{(\ell)} : x^{(\ell)}_{g, m} \in \mathcal{A}_\ell \}$. To ensure that $L$ constituent vectors, $\mathbf{x}_g^{(\ell)}$, have non-overlapping supports, candidate sets for non-zero coefficients are updated per layer as $\mathscr{M}^{(\ell)} \subseteq [M] \backslash \cup_{j=1}^{\ell-1} \mathcal{I}^{(j)}$ with $ \mathscr{M}^{(1)} \subseteq [M]$. That is, we eliminate previously selected indices to prevent duplicate usage of locations. The encoder repeatedly applies the same set of steps for $L$ layers. Thanks to such enforcement of distinct support selection, $\mathbf{x}_g$ can be represented as a superposition of $L$ sub-vectors: $\mathbf{x}_g = \sum_{\ell=1}^L \mathbf{x}_g^{(\ell)}$. The final codeword is then generated by multiplying $\mathbf{A}$ with $\mathbf{x}$: $\mathbf{c} = \mathbf{A} \mathbf{x} = \mathbf{U}_g \mathbf{x}_g$. 

It can be seen that the codeword can also be represented as a superposition of component vectors:
\begin{equation}
    \mathbf{c} = \sum_{\ell=1}^L \left( \mathbf{U}_g \mathbf{x}_g^{(\ell)} \right) = \sum_{\ell=1}^L \left( \sum_{i \in \mathcal{I}^{(\ell)}} \alpha_{\ell, \pi(i)} \mathbf{u}_{g, i} \right) = \sum_{\ell=1}^L \mathbf{c}^{(\ell)} ,
\end{equation} where $\alpha_{\ell, \pi(i)} \in \mathcal{A}_\ell$ is the alphabet value assigned to the non-zero position $i$, and $\mathbf{u}_{g, i}$ is the $i$-th column vector of $\mathbf{U}_g$, respectively. Hence, each component codeword vector $\mathbf{c}^{(\ell)}$ is a linear combination of a subset of column vectors of $\mathbf{U}_g$.
\subsection{Approximate MAP Decoder}
Let us define the entire set of possible vectors that $\mathbf{x}$ can take on as 
\begin{equation}
    \mathscr{X} := \left\{ \mathbf{x} \in \mathbb{C}^N : \mathbf{x} \in \mathcal{A}^N , \| \mathbf{x} \|_0 = K, \sum_{g=1}^G \mathbbm{1}_{\{ \| \mathbf{x}_g \|_0 \ne 0 \}} = 1 \right\} , \label{total message vector set}
\end{equation} where $\mathcal{A} := \{ 0 \} \bigcup \left( \cup_{\ell=1}^L \mathcal{A}_\ell \right)$. As the selected block index is not available at the receiver side, we set up a hypothesis that $\mathbf{U}_g$ participates in encoding and denote it by $\mathscr{H}_g$:
\begin{equation}
    \mathscr{H}_g : \mathbf{c} = \mathbf{A} \mathbf{x} = \mathbf{U}_g \mathbf{x}_g.
\end{equation} Then, we can define the sub-message vector set with respect to each hypothesis:
\begin{equation}
    \mathscr{X}_g := \left\{ \mathbf{x} \in \mathbb{C}^M : \mathbf{x}_g \in \mathcal{A}^M , \| \mathbf{x}_g \|_0 = K \right\}. \label{sub-message vector set}
\end{equation} Now, the MAP decoding problem can be formulated and dissected as follows:
\begin{align*}
    \hat{\mathbf{x}}^\text{MAP} &= \underset{\mathbf{x} \in \mathscr{X}}{\arg \max} \, \mathbb{P} ( \mathbf{x} | \mathbf{y} ) = \underset{\mathbf{x}_g \in \mathscr{X}_g , g \in [G]}{\arg \max} \, \mathbb{P} ( \mathbf{x}_g , \mathscr{H}_g | \mathbf{y} ) \\
    &= \underset{\mathbf{x}_g \in \mathscr{X}_g , g \in [G]}{\arg \max} \, \mathbb{P} ( \mathbf{x}_g | \mathbf{y} , \mathscr{H}_g ) \mathbb{P} ( \mathscr{H}_g | \mathbf{y} ) . \numberthis{\label{original MAP problem}}
\end{align*} The original two-stage MAP decoder solves the two sub-MAP tasks separately. In the first stage, the decoder finds the most likely message vector segment within $\mathscr{X}_g$ for each hypothesis:
\begin{equation}
    \hat{\mathbf{x}}_g^\text{MAP} = \underset{\hat{\mathbf{x}}_g \in \mathscr{X}_g}{\arg \max} \, \mathbb{P} ( \hat{\mathbf{x}}_g | \mathbf{y} , \mathscr{H}_g ). \label{two-stage MAP_first stage}
\end{equation} Under $\mathscr{H}_g$, it is postulated that few meaningful elements of $\mathbf{x}$, all placed in $\mathbf{x}_g$, are distributed by $\mathbf{U}_g$ and become correlated with each other. Since each sub-dictionary matrix is unitary, the effects of $\mathbf{U}_g$ can be readily removed from $\mathbf{y}$. Let us denote the decorrelated output by $\mathbf{y}_g$:
\begin{equation}
    \mathbf{y}_g = \mathbf{U}_g^\dagger \mathbf{y} = \mathbf{U}_g^\dagger ( \mathbf{U}_g \mathbf{x}_g + \mathbf{v}) = \mathbf{x}_g + \Tilde{\mathbf{v}}_g ,
\end{equation} where $\Tilde{\mathbf{v}}_g := \mathbf{U}_g^\dagger \mathbf{v}$. Note that $\Tilde{\mathbf{v}}_g \overset{(d)}{=} \mathbf{v}$ due to rotational invariance of the Gaussian distribution. Then, the first-stage decoding in \eqref{two-stage MAP_first stage} can be re-formulated as 
\begin{equation}
    \hat{\mathbf{x}}_g^\text{MAP} = \underset{\hat{\mathbf{x}}_g \in \mathscr{X}_g}{\arg \max} \, \mathbb{P} ( \hat{\mathbf{x}}_g | \mathbf{y} , \mathscr{H}_g ) = \underset{\hat{\mathbf{x}}_g \in \mathscr{X}_g}{\arg \max} \, \mathbb{P} ( \hat{\mathbf{x}}_g | \mathbf{y}_g ) . \label{two-stage MAP_first stage_v2}
\end{equation} Taking the log of the a posteriori probability (APP) in \eqref{two-stage MAP_first stage_v2} yields
\begin{align*}
    & \log \mathbb{P} ( \hat{\mathbf{x}}_g | \mathbf{y}_g ) \\
    &= \sum_{\ell=1}^L \log \mathbb{P} ( \hat{\mathbf{x}}_g^{(\ell)} | \mathbf{y}_g , \hat{\mathbf{x}}_g^{(\ell - 1)} , \dots, \hat{\mathbf{x}}_g^{(2)} , \hat{\mathbf{x}}_g^{(1)} )  \\
    & \overset{(a)}{=} \sum_{\ell=1}^L \log \mathbb{P} ( \hat{\mathbf{x}}_g^{(\ell)} | \mathbf{y}_g , \hat{\mathcal{I}}_g^{(\ell-1)} , \dots, \hat{\mathcal{I}}_g^{(2)} , \hat{\mathcal{I}}_g^{(1)}) \\
    &= \sum_{\ell=1}^L \Biggl[ \kappa \cdot \sum_{m \in \hat{\mathscr{M}}_g^{(\ell)}} \Biggl( \log \frac{\mathbb{P}(y_{g,m}|\hat{x}_{g,m}^{(\ell)}) \mathbb{P}( \hat{x}_{g,m}^{(\ell)}  )}{\mathbb{P}(y_{g,m})} \\
    & \quad \quad \quad \quad \quad \quad \quad \quad \quad \quad \quad \quad \quad \quad  \cdot \mathbbm{1}_{\{ \| \hat{\mathbf{x}}_g \|_0 = K_\ell \} } \Biggr) \Biggr], \numberthis{\label{two-stage MAP_first stage_joint log APP}}
\end{align*} where $\hat{\mathcal{I}}_g^{(\ell)}$ is the non-zero support estimate of $\hat{\mathbf{x}}_g^{(\ell)}$, and $\kappa$ is the normalization term. An estimate of the $\ell$-th layer's candidate set is updated by incorporating prior support estimates: $\hat{\mathscr{M}}_g^{(\ell)} \subseteq [M] \backslash \cup_{j=1}^{\ell-1} \hat{\mathcal{I}}_g^{(j)}$ with $| \hat{\mathscr{M}}_g^{(\ell)} | = | \mathscr{M}^{(\ell)} |$. The equality $(a)$ in \eqref{two-stage MAP_first stage_joint log APP} comes from that $\{ \hat{\mathcal{I}}_g^{(j)} \}_{j=1}^{\ell-1}$ provide sufficient information to decode $\hat{\mathbf{x}}_g^{(\ell)}$ as $\{ \hat{\mathbf{x}}_g^{(j)} \}_{j=1}^{\ell-1}$.

By comparing $G$ candidates, $\{ \hat{\mathbf{x}}_g \}_{g \in G}$, the decoder performs hypothesis testing and identifies the block index:
\begin{align*}
    \hat{g} &= \underset{g \in [G]}{\arg \max} \, \mathbb{P} ( \mathscr{H}_g | \mathbf{y} ) = \underset{g \in [G]}{\arg \max} \, \mathbb{P} ( \mathbf{c} = \mathbf{U}_g \hat{\mathbf{x}}_g^\text{MAP} | \mathbf{y} ) \\
    &= \underset{g \in [G]}{\arg \min} \, \| \mathbf{y} - \mathbf{U}_g \hat{\mathbf{x}}_g^\text{MAP} \|_2 . \numberthis{\label{two-stage MAP_second stage}}
\end{align*} The final estimate can be obtained by padding $\hat{\mathbf{x}}_{\hat{g}}^\text{MAP}$ with $M \cdot (G - 1)$ zero's, i.e., $\hat{\mathbf{x}}^\text{MAP} = \left[ \mathbf{0}_M^\mathsf{T} \cdots  {\hat{\mathbf{x}}_{\hat{g}}^\text{MAP}}\tran  \cdots \mathbf{0}_M^\mathsf{T} \right]^\mathsf{T} $.

\section{Joint Equalization and Decoding Algorithm}
This section presents a novel joint equalization-and-decoding algorithm for BOSS codes in Rayleigh multi-path fading channels with OFDM transmission. The two-stage MAP decoder leans on mutual orthogonality between component vectors of codewords, so it can no longer be used when the orthogonal property is destroyed by fading effects.
\subsection{Limits}
We first explicate why the original two-stage MAP decoding algorithm cannot be directly applied in a multi-path fading scenario. Suppose that the decoder attempts to decorrelate the elements of a putative codeword $\mathbf{x}_g$ under $\mathscr{H}_g$. The resultant vector $\mathbf{y}_g$ is given by
\begin{equation}
    \mathbf{y}_g = \mathbf{U}_g^\dagger \mathbf{y} = \mathbf{U}_g^\dagger ( \mathbf{\Lambda} \mathbf{U}_g \mathbf{x}_g + \mathbf{v} ) = \Check{\mathbf{x}}_g + \Check{\mathbf{v}}_g , \label{SISO decorrelated received vector}
\end{equation} where $\Check{\mathbf{x}}_g = \mathbf{U}_g^\dagger \mathbf{\Lambda} \mathbf{U}_g \mathbf{x}_g$, and $\Check{\mathbf{v}}_g = \mathbf{U}_g \mathbf{v}$. For ease of explanation, we suppose that the encoder select the first $L$ indices in a consecutive order through $L$ layers and that every alphabet be a singleton. In other words, $\mathcal{I}^{(\ell)} = \{ \ell \}$ and $\mathcal{A}_\ell = \{ \alpha_\ell \}$ for every $\ell \in [L]$. Then, a corresponding codeword carried by OFDM sub-carriers can be written as 
\begin{align*}
    & \mathbf{\Lambda} \mathbf{U}_g \mathbf{x}_g \\
    &\!=\! \begin{bmatrix}
        \lambda_1 \left( \sum_{\ell=1}^L \alpha_\ell ( \mathbf{U}_g )_{1, \ell} \right) & \! \cdots \! & \lambda_M \left( \sum_{\ell=1}^L \alpha_\ell ( \mathbf{U}_g )_{M, \ell} \right) 
    \end{bmatrix}^\mathsf{T} . \numberthis
\end{align*} Because of $\mathbf{\Lambda}$, $\Check{\mathbf{x}}_g$ exhibits a structural division, where elements with indices up to $L$ can be categorized separately from the remaining elements indexed from $L + 1$ to $M$:
\begin{align*}
    & \Check{x}_{g, i} = \\
    & \begin{cases}
        \begin{aligned}
            & \alpha_i \Biggl( \underbrace{\sum_{m=1}^M \lambda_m \| ( \mathbf{U}_g )_{m, i} \|_2^2 }_{\ne 1}  \Biggr)  \\
            & + \sum_{\ell \in [L] \backslash \{ i \}} \alpha_j \cdot \Biggl( \underbrace{\sum_{m=1}^M \lambda_m ( \mathbf{U}_g )_{m, i}^* ( \mathbf{U}_g )_{m, \ell}}_{\ne 0} \Biggr) , && \text{if } 1 \leq i \leq L \\
            & \sum_{\ell \in [L]} \alpha_\ell \cdot \Biggl( \underbrace{\sum_{m=1}^M \lambda_m ( \mathbf{U}_g )_{m, i}^* ( \mathbf{U}_g )_{m, \ell}}_{\ne 0} \Biggr) , && \text{else. }  
        \end{aligned} 
    \end{cases} \numberthis
\end{align*} In this scenario, $\Check{\mathbf{x}}_g$ losses sparsity, and correlations persist among its entries. Consequently, computing the MAP metric for each element becomes challenging, as it necessitates the consideration of the effects of other layers. This complexity arises because previous support estimates lack sufficient information to reconstruct the segment vectors for each layer.

\subsection{MMSE-A-MAP Decoder}
\textbf{1st stage MMSE equalization:} Let us introduce a new notation, $\mathbf{c}_g = \mathbf{U}_g \mathbf{x}_g$, for a putative codeword under $\mathscr{H}_g$. We, then, postulate that for all $g \in [G]$, elements of $\mathbf{c}_g$ are independent and identically distributed zero-mean complex Gaussian with variance $\sigma_{\mathbf{c}_g}^2$, i.e., $\mathbf{c}_g \sim \mathcal{CN}( \mathbf{0}_M , \Tilde{\mathbf{K}}_{\mathbf{c}_g \mathbf{c}_g})$ with $\Tilde{\mathbf{K}}_{\mathbf{c}_g \mathbf{c}_g} = \sigma_{\mathbf{c}_g}^2 \mathbf{I}_M$. The value of $\sigma_{\mathbf{c}_g}^2$ is calculated as 
\begin{equation}
    \sigma_{\mathbf{c}_g}^2 = \frac{\text{tr} ( \mathbf{U}_g \mathbf{K}_{\mathbf{x}_g \mathbf{x}_g } \mathbf{U}_g^\dagger )}{M}, \label{codeword variance}
\end{equation} where $\mathbf{K}_{\mathbf{x}_g \mathbf{x}_g} \in \mathbb{C}^{M \times M}$ is the sample covariance matrix of $\mathbf{x}_g$. Since the sparsity level $K_\ell$, constellation $\mathcal{A}_\ell$ per layer, and candidate set $\mathscr{M}^{(\ell)}$ are fixed a priori, $\mathbf{K}_{\mathbf{x}_g \mathbf{x}_g}$ can be readily computed offline by taking account of every possible selection of the non-zero coefficients. Moreover, the trace is invariant under cyclic permutations, so $\sigma_{\mathbf{c}_g}^2$ in \eqref{codeword variance} is reduced to $\frac{\text{tr}(\mathbf{K}_{\mathbf{x}_g \mathbf{x}_g})}{M}$ and becomes independent of $\mathbf{U}_g$ . We, hence, drop the block index notation in $\Tilde{\mathbf{K}}_{\mathbf{c}_g \mathbf{c}_g}$ and $\sigma_{\mathbf{c}_g}^2$, and obtain an MMSE equalizer as follows:
\begin{align*}
    &\mathbf{W}^\text{MMSE} \\
    &:= \mathbf{K}_{\mathbf{c}_g \mathbf{y}} \mathbf{K}_{\mathbf{y} \mathbf{y}}^{-1} \\
    &= \mathbb{E} \{ \mathbf{c}_g ( \mathbf{\Lambda} \mathbf{c}_g + \mathbf{v} )^\dagger \} \left( \mathbb{E} \{ ( \mathbf{\Lambda} \mathbf{c}_g + \mathbf{v} ) ( \mathbf{\Lambda} \mathbf{c}_g + \mathbf{v} )^\dagger \} \right)^{-1} \\
    &= \Tilde{\mathbf{K}}_{\mathbf{c} \mathbf{c}} \mathbf{\Lambda}^\dagger \left( \mathbf{\Lambda} \Tilde{\mathbf{K}}_{\mathbf{c} \mathbf{c}} \mathbf{\Lambda}^\dagger + \sigma_v^2 \mathbf{I}_M \right)^{-1} \\
    &= \text{diag} \left( \frac{\lambda_1^* \sigma_{\mathbf{c}}^2}{| \lambda_1 |^2 \sigma_{\mathbf{c}}^2 + \sigma_v^2 } , \frac{\lambda_2^* \sigma_{\mathbf{c}}^2}{| \lambda_2 |^2 \sigma_{\mathbf{c}}^2 + \sigma_v^2 } , \dots , \frac{\lambda_M^* \sigma_{\mathbf{c}}^2}{| \lambda_M |^2 \sigma_{\mathbf{c}}^2 + \sigma_v^2 } \right) . \numberthis{\label{MMSE equalizer}}
\end{align*} It is obvious that the resultant MMSE filter is independent of the dictionary matrix, so the same filter can be used within the coherent period across different hypotheses.

\textbf{2nd stage A-MAP: } By applying decorrelation to the MMSE equalizer output,  \eqref{SISO decorrelated received vector} can be re-written as
\begin{equation}
    \mathbf{y}_g^\text{MMSE} = \mathbf{U}_g^\dagger \mathbf{W}^\text{MMSE} \mathbf{y} = \Tilde{\mathbf{x}}_g + \Tilde{\mathbf{v}}_g , \label{MMSE-A-MAP decorrelation output}
\end{equation} where $\Tilde{\mathbf{x}}_g = \mathbf{U}_g^\dagger \mathbf{W}^\text{MMSE} \mathbf{\Lambda} \mathbf{U}_g \mathbf{x}_g$ and $\Tilde{\mathbf{v}}_g = \mathbf{U}_g^\dagger \mathbf{W}^\text{MMSE} \mathbf{v}$ denote an estimate of $\mathbf{x}_g$ whose orthogonality is partially restored, and effective noise, respectively. Now, the decoder performs approximate MAP (A-MAP) decoding:
\begin{equation}
    \hat{\mathbf{x}}_g^\text{A-MAP} = \underset{\hat{\mathbf{x}}_g \in \mathscr{X}_g}{\arg \max} \, \log \mathbb{P} ( \Tilde{\mathbf{x}}_g | \mathbf {y}_g^\text{MMSE} ). \label{MMSE-A-MAP 2nd stage A-MAP}
\end{equation} By invoking the chain rule, a log APP in \eqref{MMSE-A-MAP 2nd stage A-MAP} can be represented as the sum of $L$ log likelihoods:
\begin{multline}
    \log \mathbb{P} ( \Tilde{\mathbf{x}}_g | \mathbf{y}_g^\text{MMSE} ) \\
    = \sum_{\ell=1}^L \log \mathbb{P} ( \Tilde{\mathbf{x}}_g^{(\ell)} | \mathbf{y}_g^\text{MMSE}, \Tilde{\mathbf{x}}_g^{(\ell - 1)}, \dots, \Tilde{\mathbf{x}}_g^{(2)} , \Tilde{\mathbf{x}}_g^{(1)}), \label{MMSE-A-MAP 2nd stage log APP}
\end{multline} where $\Tilde{\mathbf{x}}_g^{(\ell)} = \mathbf{U}_g^\dagger \mathbf{W}^\text{MMSE} \mathbf{\Lambda} \mathbf{U}_g \mathbf{x}_g^{(\ell)}$.

Since $\mathbf{U}_g^\dagger \mathbf{W}^\text{MMSE}$ is no longer unitary, elements of $\Tilde{\mathbf{v}}_g$ are statistically correlated with each other. Through numerous simulation runs, it was consistently observed that the sample covariance matrix of $\Tilde{\mathbf{v}}_g$ tends to be diagonally dominant. This observation suggests that the correlation within $\Tilde{\mathbf{v}}_g$ is relatively weak when juxtaposed with the variances of individual elements. Building upon this empirical evidence, we articulate our second assumption, positing that the elements of $\Tilde{\mathbf{v}}_g$ are independent zero-mean complex Gaussian with varying variances of
\begin{equation}
    \sigma_{\Tilde{v}_{g,m}}^2 = \sigma_v^2 \cdot \left( \sum_{j=1}^M \| ( \mathbf{U}_g )_{j, m}^* w_j^\text{MMSE} \|_2^2 \right), \quad \forall m \in [M], \label{MMSE-A-MAP effective noise variance}
\end{equation} where $w_j^\text{MMSE}$ is the $j$-th diagonal element of $\mathbf{W}^\text{MMSE}$. Under this assumption, elements of $\mathbf{y}_g^\text{MMSE}$ become statistically independent, conditioned upon $\Tilde{\mathbf{x}}_g$. The $\ell$-th layer's log APP in \eqref{MMSE-A-MAP 2nd stage log APP} can be re-written and and factorized as follows:
\begin{align*}
    & \log \mathbb{P} ( \Tilde{\mathbf{x}}_g^{(\ell)} | \mathbf{y}_g^\text{MMSE} , \Tilde{\mathbf{x}}_g^{(\ell-1)} , \dots , \Tilde{\mathbf{x}}_g^{(2)} , \Tilde{\mathbf{x}}_g^{(1)} ) \\
    &= \log \mathbb{P} ( \Tilde{\mathbf{x}}_g^{(\ell)} | \mathbf{y}_g^\text{MMSE} , \hat{\mathbf{x}}_g^{(\ell-1)} , \dots , \hat{\mathbf{x}}_g^{(2)} , \hat{\mathbf{x}}_g^{(1)} ) \\
    &= \kappa' \cdot \sum_{m \in \hat{\mathscr{M}}_g^{(\ell)}} \log \mathbb{P} \Bigl( \Tilde{x}_{g,m}^{(\ell)} | y_{g,m}^\text{MMSE} , \sum_{j=1}^{\ell-1} \hat{\mathbf{x}}_g^{(j)} \Bigr) \mathbbm{1}_{ \{ \| \hat{\mathbf{x}}_g^{(\ell)} \|_0 = K_\ell \}  } , \numberthis{\label{MMSE-A-MAP 2nd stage lth layer log APP}}
\end{align*} with $\kappa'$ being a normalizing term. It should be noted that the sparsity constraint in \eqref{MMSE-A-MAP 2nd stage lth layer log APP} is given in terms of $\hat{\mathbf{x}}_g^{(\ell)}$, a sub-message vector estimate to be recovered at the current layer, instead of $\Tilde{\mathbf{x}}_g^{(\ell)}$. This is because $\Tilde{\mathbf{x}}_g^{(\ell)}$ is no longer sparse, and $\Tilde{x}_{g,m}^{(\ell)} \in \left( \{ 0\} \cup \mathcal{A}_\ell \right)$. Every element of $\mathbf{y}_g^\text{MMSE}$ follows the complex Gaussian distribution of mean $\Tilde{x}_{g,m}^{(\ell)}$ and variance $\sigma_{\Tilde{v}_{g,m}}^2$ computed in \eqref{MMSE-A-MAP effective noise variance}, i.e, $y_{g,m}^\text{MMSE} \sim \mathcal{CN} ( \Tilde{x}_{g,m}^{(\ell)} , \sigma_{\Tilde{v}_{g,m}}^2)$. Then, we shall explore different possibilities of $\Tilde{\mathbf{x}}_g^{(\ell)}$ by iterating through all conceivable combinations of non-zero coefficients within $\hat{\mathbf{x}}_g^{(\ell)}$, although such an exhaustive search may incur prohibitive computational complexity. Another important deviation from decoding in the AWGN channel is that $\Tilde{\mathbf{x}}_g^{(\ell)}$ is conditioned upon supplementary information regarding message estimates of previous layers, i.e., $\hat{\mathbf{x}}_g^{(\ell-1)}, \dots, \hat{\mathbf{x}}_g^{(2)}, \hat{\mathbf{x}}_g^{(1)}$. In the absence of orthogonality in $\Tilde{\mathbf{x}}_g$, effects of support estimates from preceding $\ell - 1$ layers should be factored into the estimation of $\Tilde{\mathbf{x}}_g^{(\ell)}$. This amounts to an additional increment in complexity. Sparsity, however, comes to our rescue when the computational complexity is seemingly daunting. Taking into account that $\hat{\mathbf{x}}_g^{(\ell)}$ of each layer is $K_\ell$-sparse with $K_\ell$ being much smaller than $M$, $\Tilde{x}_{g,m}^{(\ell)}$ is a weighted linear combination of elements of the $m$-th row of $\mathbf{Q}_g = \mathbf{U}_g^\dagger \mathbf{W}^\text{MMSE} \mathbf{\Lambda} \mathbf{U}_g$. The primary focus of this paper revolves around the transmission of short data packets, so the same $\mathbf{Q}_g$ can be reused throughout the channel coherence period, thereby minimizing complexity. Finally, how to incorporate the effects of previous message vector estimates into the tasks of support identification and signal level detection will be addressed shortly.

Before formulating the decoding problem, we begin by examining an illustrative example: a two-layer BOSS code with $K_1 = K_2 = 2$, and $\mathcal{A}_1 = \{ \phi_1, \phi_2 \}$ and $\mathcal{A}_2 = \{ \psi_1 , \psi_2 \}$. In the first layer, an APP conditioned upon the event that an $m$-th element of $\hat{\mathbf{x}}_g^{(1)}$ is non-zero is given by 
\begin{align*}
    &\mathbb{P} ( y_{g,m}^\text{MMSE} | \hat{x}_{g,m}^{(1)} \in \mathcal{A}_1 )  \\
    &= \frac{1}{2} \sum_{j=1}^2 \frac{1}{\pi \sigma_{\Tilde{v}_{g,m}}^2} \exp \left(-\frac{\| y_{g,m}^\text{MMSE} - \phi_j \cdot ( \mathbf{Q}_g )_{m,m} \|_2^2}{\sigma_{\Tilde{v}_{g,m}}^2} \right) . \numberthis{\label{MMSE-A-MAP conditional APP}}
\end{align*} Invoking \eqref{MMSE-A-MAP conditional APP}, decoding in the first layer boils down to assessing and contrasting the likelihood of each index $m \in \hat{\mathscr{M}}_g^{(1)}$ belonging to $\mathcal{I}^{(1)}$ given $y_{g,m}^\text{MMSE}$:
\begin{align*}
    & \log \mathbb{P} ( m \in \mathcal{I}^{(1)} | y_{g,m}^\text{MMSE} ) \\
    &= \log \frac{ \mathbb{P} ( y_{g,m}^\text{MMSE} | \hat{x}_{g,m}^{(1)} \in \mathcal{A}_1 ) \mathbb{P} ( \hat{x}_{g,m}^{(1)} \in \mathcal{A}_1 )  }{\mathbb{P} ( y_{g,m}^\text{MMSE})} \\
    &= \log \mathbb{P} \Biggl[ \frac{1}{2} \sum_{j=1}^2 \frac{1}{\pi \sigma_{\Tilde{v}_{g,m}}^2} \\
    & \quad \quad \quad \quad \quad \quad \exp \left( - \frac{\| y_{g,m}^\text{MMSE} - \phi_j \cdot ( \mathbf{Q}_g )_{m,m} \|_2^2}{\sigma_{\Tilde{v}_{g,m}}^2} \right) p^{(1)} \Biggr] \\
    & \quad - \log \mathbb{P} \Biggl[ \frac{1}{2} \sum_{j=1}^2 \Biggl\{ \frac{1}{\pi \sigma_{\Tilde{v}_{g,m}}^2} \\
    & \quad \quad \quad \quad \quad \quad \exp \left( -\frac{\| y_{g,m}^\text{MMSE} - \phi_j \cdot ( \mathbf{Q}_g )_{m,m} \|_2^2}{\sigma_{\Tilde{v}_{g,m}}^2} \right) p^{(1)} \Biggr\} \\
    & \quad \quad \quad \quad \quad + \frac{1}{\sigma_{\Tilde{v}_{g,m}}^2} \exp \left( -\frac{\| y_{g,m}^\text{MMSE} \|_2^2}{\sigma_{\Tilde{v}_{g,m}}^2} \right) (1 - p^{(1)}) \Biggr] , \numberthis{\label{MMSE-A-MAP toy example element-wise conditional APP}}
\end{align*} where $p^{(1)} = K_1 / | \hat{\mathscr{M}}_g^{(1)} |$. The decoder sorts $\hat{\mathscr{M}}_g^{(1)}$ based on the log APP values in \eqref{MMSE-A-MAP toy example element-wise conditional APP}, and forms $\hat{\mathcal{I}}_g^{(1)} = \{ \hat{i}_{g,1}^{(1)} , \hat{i}_{g,2}^{(1)} \}$ by choosing $K_1$ indices that are most likely to be non-zero. Once $\hat{\mathcal{I}}_g^{(1)}$ is obtained, the decoder performs MAP estimation to determine the assignment of values of $\mathcal{A}_1$ to $\{ \hat{x}_{g,m}^{(1)} \}_{m \in \hat{\mathcal{I}}_g^{(1)} }$:
\begin{equation}
    \hat{x}_{g,m}^{(1)} = \underset{\phi_j \in \mathcal{A}_1}{\arg \max} \, \mathbb{P} ( \hat{x}_{g,m}^{(1)} = \phi_j | y_{g,m}^\text{MMSE} ). \label{MMSE-A-MAP toy example signal level detection}
\end{equation} Since the equalizer $\mathbf{W}^\text{MMSE}$ does not fully restore orthogonality, we shall be cautious and simultaneously consider the allocation of $\phi_1$ and $\phi_2$, taking account of their effects on each other. This concern does not arise in the AWGN channel, as the decorrelator completely nullifies any correlations between the two non-zero coefficients. The decoding task in \eqref{MMSE-A-MAP toy example signal level detection} can be re-written as follows:
\begin{align*}
    & \left\| y_{g,m}^\text{MMSE} - \bigl( \phi_1 \cdot ( \mathbf{Q}_g )_{m,m} + \phi_2 \cdot ( \mathbf{Q}_g )_{m, \hat{\mathcal{I}}_g^{(1)} \backslash \{ m \}} \bigr) \right\|_2 \\
    & \quad \overset{\hat{x}_{g,m}^{(1)} = \phi_2}{\underset{\hat{x}_{g,m}^{(1)} = \phi_1}{\gtrless}} \left\| y_{g,m}^\text{MMSE} - \bigl( \phi_2 \cdot ( \mathbf{Q}_g )_{m,m} + \phi_1 \cdot ( \mathbf{Q}_g)_{m, \hat{\mathcal{I}}_g^{(1)} \backslash {m} } \bigr) \right\|_2 , \numberthis
\end{align*} with ties broken arbitrarily. Without loss of generality, suppose that $\hat{x}_{g,\hat{i}_{g,1}^{(1)}}^{(1)} = \phi_1$ and $\hat{x}_{g,\hat{i}_{g,2}^{(1)}}^{(1)} = \phi_2$. The decoder advances to the second layer and sorts $\hat{\mathscr{M}}_g^{(2)} \subseteq [M] \backslash \hat{\mathcal{I}}_g^{(1)}$ according to the following MAP metric:
\begin{align*}
    & \log \mathbb{P} ( m \in \mathcal{I}^{(2)} | y_{g,m}^\text{MMSE}, \hat{\mathbf{x}}_g^{(1)} ) \\
    &= \log \frac{ \mathbb{P} ( y_{g,m}^\text{MMSE} | \hat{x}_{g,m}^{(2)} \in \mathcal{A}_2, \hat{\mathbf{x}}_g^{(1)} ) \mathbb{P} ( \hat{x}_{g,m}^{(2)} \in \mathcal{A}_2 | \hat{\mathbf{x}}_g^{(1)} )  }{\mathbb{P} ( y_{g,m}^\text{MMSE} | \hat{\mathbf{x}}_{g}^{(1)})} \\
    &= \log \mathbb{P} \Biggl[ \frac{1}{2} \sum_{j=1}^2 \frac{1}{\pi \sigma_{\Tilde{v}_{g,m}}^2} \exp \left( - \frac{\| y_{g,m}^\text{MMSE} - \tau_{j, m} \|_2^2}{\sigma_{\Tilde{v}_{g,m}}^2} \right) p^{(2)} \Biggr] \\
    & \quad - \log \mathbb{P} \Biggl[ \frac{1}{2} \sum_{j=1}^2 \Biggl\{ \frac{1}{\pi \sigma_{\Tilde{v}_{g,m}}^2} \exp \left( -\frac{\| y_{g,m}^\text{MMSE} - \tau_{j, m} \|_2^2}{\sigma_{\Tilde{v}_{g,m}}^2} \right) p^{(2)} \Biggr\} \\
    & \quad \quad \quad \quad \quad + \frac{1}{\sigma_{\Tilde{v}_{g,m}}^2} \exp \left( -\frac{\| y_{g,m}^\text{MMSE} - \gamma_m \|_2^2}{\sigma_{\Tilde{v}_{g,m}}^2} \right) (1 - p^{(2)}) \Biggr] , \numberthis
\end{align*} where
\begin{align}
    \gamma_m &= \phi_1 \cdot ( \mathbf{Q}_g )_{m, \hat{i}_{g,1}^{(1)}} + \phi_2 \cdot ( \mathbf{Q}_g )_{m, \hat{i}_{g,2}^{(1)}} , \label{MMSE-A-MAP toy example gamma} \\
    \tau_{j, m} &= \psi_j \cdot ( \mathbf{Q}_g )_{m,m} + \gamma_m , \text{ and} \label{MMSE-A-MAP toy example tau} \\
    p^{(2)} &= \frac{K_2}{| \hat{\mathscr{M}}_g^{(2)} |} .
\end{align} It can be seen that the effects of $\hat{\mathbf{x}}_g^{(1)}$ onto computing $\Tilde{x}_{g,m}^{(2)}$ are reflected in $\gamma_m$. The decoder identifies $\hat{\mathcal{I}}_g^{(2)}$, comprising the $K_2$ most probable indices from $\hat{\mathscr{M}}_g^{(2)}$. For $m \in \hat{\mathcal{I}}_g^{(2)}$, the decoder solves the following signal level detection problem:
\begin{align*}
    \left\| y_{g,m}^\text{MMSE} - (\Psi_{12} + \gamma_m ) \right\|_2 \overset{\hat{x}_{g,m}^{(2)}=\psi_2}{\underset{\hat{x}_{g,m}^{(2)} = \psi_1}{\gtrless}} \left\| y_{g,m}^\text{MMSE} - (\Psi_{21} + \gamma_m) \right\|_2 ,
\end{align*} where $\Psi_{12} = \psi_1 \cdot ( \mathbf{Q}_g )_{m,m} + \psi_2 \cdot ( \mathbf{Q}_g )_{m, \hat{\mathcal{I}}_g^{(2)} \backslash \{ m \} }$ and $\Psi_{21} = \psi_2 \cdot ( \mathbf{Q}_g )_{m,m} + \psi_1 \cdot ( \mathbf{Q}_g )_{m, \hat{\mathcal{I}}_g^{(2)} \backslash \{ m \} }$, and $\gamma_m$ is given in \eqref{MMSE-A-MAP toy example gamma}. Again, ties are to be broken arbitrarily.

We now return to the general case and revisit the first layer. The decoder evaluates the likelihood of $m \in \hat{\mathscr{M}}_g^{(1)}$ being a member of $\mathcal{I}^{(1)}$:
\begin{align*}
    & \log \mathbb{P} ( m \in \mathcal{I}^{(1)} | y_{g,m}^\text{MMSE} ) \\
    &= \log \mathbb{P} \Biggl[ \frac{1}{K_\ell} \sum_{j=1}^{K_\ell} \frac{1}{\pi \sigma_{\Tilde{v}_{g,m}}^2} \\
    & \quad \quad \quad \quad \quad \quad \exp \left( - \frac{\| y_{g,m}^\text{MMSE} - \alpha_{1,j} \cdot ( \mathbf{Q}_g )_{m,m} \|_2^2}{\sigma_{\Tilde{v}_{g,m}}^2} \right) p^{(1)} \Biggr] \\
    & \quad - \log \mathbb{P} \Biggl[ \frac{1}{K_\ell} \sum_{j=1}^{K_\ell} \Biggl\{ \frac{1}{\pi \sigma_{\Tilde{v}_{g,m}}^2} \\
    & \quad \quad \quad \quad \quad \quad \exp \left( -\frac{\| y_{g,m}^\text{MMSE} - \alpha_{1,j} \cdot ( \mathbf{Q}_g )_{m,m} \|_2^2}{\sigma_{\Tilde{v}_{g,m}}^2} \right) p^{(1)} \Biggr\} \\
    & \quad \quad \quad \quad \quad + \frac{1}{\sigma_{\Tilde{v}_{g,m}}^2} \exp \left( -\frac{\| y_{g,m}^\text{MMSE} \|_2^2}{\sigma_{\Tilde{v}_{g,m}}^2} \right) (1 - p^{(1)}) \Biggr] .
\end{align*} Let us denote by $\mathscr{P}^{(1)}$ a super set of $K_1 !$ different ways to arrange elements of $\mathcal{A}_1$. For example, if $\mathcal{A}_1 = \{ \phi_1 , \phi_2 \}$, then $\mathscr{P}^{(1)} = \{ \mathscr{P}^{(1)}_1 , \mathscr{P}_2^{(1)} \} = \left\{ \{ \phi_1, \phi_2 \} , \{ \phi_2, \phi_1 \} \right\}$. Note that $\mathscr{P}^{(\ell)}$ is not dependent on $\mathscr{H}_g$ since the sparsity $K_\ell$ and constellation $\mathcal{A}_\ell$ per layer are fixed a priori. The decoder seeks an optimal way $\mathscr{P}_{j^*}^{(1)}$ to allocate values of $\mathcal{A}_1$ to $\{ \hat{x}_{g,m}^{(1)} \}_{m \in \hat{\mathcal{I}}_g^{(1)} }$:
\begin{equation}
    j^* = \underset{j \in [K_1 ! ]}{\arg \min} \, \sum_{k \in [K_1]} \| y_{g, \hat{i}_{g, k}^{(1)}}^\text{MMSE} - \Gamma_{j, k}^{(1)} \|_2 ,  
\end{equation} where $\Gamma_{j, k}^{(1)}$ is an approximate mean of $y_{g, \hat{i}_{g, k}^{(1)}}^\text{MMSE}$ when values of $\mathcal{A}_1$ are assigned according to $\mathscr{P}^{(1)}_j$. That is,
\begin{equation}
    \Gamma_{j, k}^{(1)} = \sum_{m \in [K_1]} \mathdutchcal{p}_{j,m}^{(1)} \cdot ( \mathbf{Q}_g )_{\hat{i}_{g,k}^{(1)} , \hat{i}_{g,m, \phantom{k}}^{(1)} } ,
\end{equation} where $\mathdutchcal{p}_{1,m}^{(1)}$ is the $m$-th value of $\mathscr{P}_j^{(1)}$. As a result, the natural order of non-zero coefficients in $\hat{\mathbf{x}}_g^{(1)}$ is consistent with the order of alphabets in $\mathscr{P}_{j^*}^{(1)}$, i.e., $ \mathscr{P}_{j^*}^{(1)} = \left\{ \hat{x}_{g, \hat{i}_{g, 1}^{(1)}}^{(1)} , \dots , \hat{x}_{g, \hat{i}_{g, K_1}^{(1)}}^{(1)} \right\}$. Note that we omit the block index $g$ in the optimal set index $j^*$ although it varies by $\mathscr{H}_g$. This is because it is only required in subsequent layers to estimate $\hat{\mathbf{x}}_g$, and passed onto neither the next stage nor decoding process under other hypotheses. The support identification task in the $\ell$-th layer is to evaluate the metric in \eqref{MMSE-A-MAP 2nd stage element-wise MAP} for every index $m \in \hat{\mathscr{M}}_g^{(\ell)}$, where
\begin{figure*}
    \centering
    \normalsize
    \begin{align*}
        & \log \mathbb{P} ( m \in \mathcal{I}^{(\ell)} | y_{g,m}^\text{MMSE} , \hat{\mathbf{x}}_g^{(\ell - 1)} , \dots , \hat{\mathbf{x}}_g^{(2)} , \hat{\mathbf{x}}_g^{(1)} ) \\
        &= \log  \frac{ \frac{1}{K_\ell} \sum_{j=1}^{K_\ell} \frac{1}{\pi \sigma_{\Tilde{v}_{g,m}}^2} \exp \left( - \frac{ \| y_{g,m}^\text{MMSE} - \tau_{m, j}^{(\ell)} \|_2^2  }{\sigma_{\Tilde{v}_{g,m}}^2} \right) p^{(\ell)}    }{\frac{1}{K_\ell} \sum_{j=1}^{K_\ell} \frac{1}{\pi \sigma_{\Tilde{v}_{g,m}}^2} \exp \left( - \frac{ \| y_{g,m}^\text{MMSE} - \tau_{m, j}^{(\ell)} \|_2^2  }{\sigma_{\Tilde{v}_{g,m}}^2} \right) p^{(\ell)}  + \frac{1}{\pi \sigma_{\Tilde{v}_{g,m}}^2} \exp \left( -\frac{\| y_{g,m}^\text{MMSE} - \gamma_m^{(\ell-1)} \|_2^2}{\sigma_{\Tilde{v}_{g,m}}^2} \right) (1 - p^{(\ell)} )  } \numberthis{\label{MMSE-A-MAP 2nd stage element-wise MAP}}
    \end{align*}
    \hrulefill
    \vspace{-2mm}
\end{figure*}
\begin{align*}
    \gamma_m^{(\ell-1)} &= \sum_{t=1}^{\ell - 1} \left( \sum_{k=1}^{K_t} \hat{x}^{(t)}_{g, \hat{i}_{g,k}^{(t)}  } \cdot ( \mathbf{Q}_g )_{m, \hat{i}_{g,k}^{(t)} } \right) \text{ and} \numberthis{\label{MMSE-A-MAP 2nd stage gamma}} \\
    \tau_{m, j}^{(\ell)} &= \alpha_{\ell, j} \cdot ( \mathbf{Q}_g )_{m, m} + \gamma_m^{(\ell - 1)} \numberthis{\label{MMSE-A-MAP 2nd stage tau}} .
\end{align*} Then, the signal detection task is to find the optimal arrangement $\mathscr{P}^{(\ell)}_{j^*}$ utilizing $\hat{\mathcal{I}}_g^{(\ell)}$ and $\hat{\mathbf{x}}_g^{(1)}, \hat{\mathbf{x}}_g^{(2)} , \dots , \hat{\mathbf{x}}_g^{(\ell - 1)}$:
\begin{equation}
    j^* = \underset{j \in [K_\ell ! ]}{\arg \min} \sum_{k \in [K_\ell]} \| y_{g, \hat{i}_{g, k}^{(\ell)} }^\text{MMSE} - \Gamma_{j, k}^{(\ell)} \|_2 ,
\end{equation} where 
\begin{equation}
    \Gamma_{j, k}^{(\ell)} = \left( \sum_{m \in [ K_\ell ]} \mathdutchcal{p}_{j, m}^{(\ell)} \cdot ( \mathbf{Q}_g )_{ \hat{i}_{g, k}^{(\ell)}, \hat{i}_{g, m, \phantom{k}}^{(\ell)}  } \right) + \gamma_{\hat{i}_{g, k}^{(\ell)}}^{(\ell - 1)}  ,
\end{equation} with $\gamma_{\hat{i}^{(\ell)}_{g, k}}^{(\ell - 1)} $ defined in \eqref{MMSE-A-MAP 2nd stage gamma}. After $L$ iteration runs, the decoder obtains $\hat{\mathbf{x}}_g^\text{A-MAP}$.

\textbf{3rd stage    hypothesis testing: } In the final stage, the decoder determines which sub-dictionary matrix has participated in the encoding process. This can be accomplished through hypothesis testing:
\begin{equation}
    \hat{g} = \underset{g \in [G]}{\arg \max} \, \mathbb{P} ( \mathscr{H}_g | \mathbf{y} ) = \underset{g \in [G]}{\arg \min} \, \| \mathbf{y} - \mathbf{\Lambda} \mathbf{U}_g \hat{\mathbf{x}}_g^\text{A-MAP} \|_2 . \label{MMSE-A-MAP third stage hypothesis testing}
\end{equation} The equality in \eqref{MMSE-A-MAP third stage hypothesis testing} is due to $\mathbf{v} \sim \mathcal{CN}( \mathbf{0}_M, \sigma_v^2 \mathbf{I}_M )$ and $g \sim \mathcal{U} \{1, G\}$. The detailed steps of the proposed MMSE-A-MAP algorithm are provided in Algorithm I.

\begin{algorithm}
    \caption{\textbf{Algorithm I: } MMSE-A-MAP Decoding}
    \begin{algorithmic}[1]
        \Require received vector $\mathbf{y}$, channel marix $\mathbf{\Lambda}$, and diagonal matrix $\Tilde{\mathbf{K}}_{\mathbf{c} \mathbf{c}} = \sigma_{\mathbf{c}}^2 \mathbf{I}_M$ with $\sigma_{\mathbf{c}}^2 = \frac{\text{tr}( \mathbf{K}_{\mathbf{x}_g \mathbf{x}_g} )}{M}$
        \Ensure sparse message vector $\hat{\mathbf{x}}^\text{MMSE-A-MAP} \in \mathcal{A}^N$ with $\| \hat{\mathbf{x}}^\text{MMSE-A-MAP} \|_0 = K$
        \State Compute $\mathbf{W}^\text{MMSE} = \Tilde{\mathbf{K}}_{\mathbf{c} \mathbf{c}} \mathbf{\Lambda} ( \mathbf{\Lambda} \Tilde{\mathbf{K}}_{\mathbf{c} \mathbf{c}} \mathbf{\Lambda}^\dagger + \sigma_v^2 \mathbf{I}_M )^{-1}$
        \For{$g = 1, \dots, G$}
            \State Compute $\mathbf{y}_g^\text{MMSE} = \mathbf{U}_g^\dagger \mathbf{W}^\text{MMSE} \mathbf{y}$
            \State Calculate effective noise variance values $\sigma_{\Tilde{v}_{g,m}}^2$ \eqref{MMSE-A-MAP effective noise variance}
            \For{$\ell = 1, \dots, L$}
                \State Configure $\hat{\mathscr{M}}_g^{(\ell)} \subseteq [M] \backslash \cup_{\j=1}^{\ell-1} \hat{\mathcal{I}}_g^{(j)}$
                \LComment{$\hat{\mathscr{M}}_g^{(1)} = \mathscr{M}^{(1)}$}
                \State Calculate the metric \eqref{MMSE-A-MAP 2nd stage element-wise MAP} for every $m \in \hat{\mathscr{M}}_g^{(\ell)}$
                \State Identify $K_\ell$ indices of the largest metric values and construct $\hat{\mathcal{I}}_g^{(\ell)}$
                \State Find the optimal arrangement $\mathscr{P}_{j^*}^{(\ell)}$
                \State Obtain $\hat{\mathbf{x}}_g^{\text{A-MAP}, (\ell)} $
            \EndFor
            \State Acquire $\hat{\mathbf{x}}_g^{\text{A-MAP}} = \sum_{\ell=1}^L \hat{\mathbf{x}}_g^{\text{A-MAP}, (\ell)}$
        \EndFor
        \State Perform hypothesis testing and identify the most-likely candidate $\hat{\mathbf{x}}_{\hat{g}}^\text{A-MAP}$
        \State \Return $\hat{\mathbf{x}}^\text{MMSE-A-MAP} = \begin{bmatrix}
            \mathbf{0}_M^\mathsf{T} & \cdots & {\hat{\mathbf{x}}_{\hat{g}}^\text{A-MAP}}\tran & \cdots & \mathbf{0}_M^\mathsf{T}
        \end{bmatrix}$
    \end{algorithmic}
\end{algorithm}

\subsection{Complexity Analysis}
Since both $\Tilde{\mathbf{K}}_{ \mathbf{c} \mathbf{c} }$ and $\mathbf{\Lambda}$ are diagonal matrices, computing $\mathbf{W}^\text{MMSE}$ in \eqref{MMSE equalizer} is straightforward. Moreover, it can be reused within the channel coherent period. Thus, setting aside the corresponding computation, we provide a detailed complexity analysis of the second and third stage. Under each hypothesis $\mathscr{H}_g$, the A-MAP stage consists of the following operations:
\begin{itemize}
    \item \textbf{Decorrelation:} The obtained MMSE equalizer is diagonal, and the sub-dictionary matrix $\mathbf{U}_g$ supports fast unitary transformation. Hence, $\mathbf{y}_g^\text{MMSE}$ in \eqref{MMSE-A-MAP decorrelation output} can be obtained with a complexity of $\mathcal{O}(M \log M + M)$.
    \item \textbf{Effective noise variance computation:} For every $m \in [M]$, $\sigma_{\Tilde{v}_{g,m}}^2$ in \eqref{MMSE-A-MAP effective noise variance} should be computed, which is a simple summation and requires $\mathcal{O} (M)$ complexity.
    \item \textbf{Element-wise A-MAP metric calculation:} At each $\ell$-th layer, the log likelihood metric in \eqref{MMSE-A-MAP 2nd stage element-wise MAP} should be calculated for every $m \in \hat{\mathscr{M}}_g^{(\ell)}$. Including the complexity of $\mathcal{O}(\sum_{t=1}^{\ell-1} K_t )$ required to compute $\gamma_m^{(\ell - 1)}$ and $\tau_{m, j}^{(\ell)}$, the total complexity is $\mathcal{O} ( | \mathscr{M}^{(\ell)} | \cdot K_\ell + \sum_{t=1}^{\ell - 1} K_t ) $.
    \item \textbf{Support estimate construction:} Sorting $| \mathscr{M}^{(\ell)} |$ metrics and identifying $K_\ell$ largest entries can be done in $\mathcal{O} ( | \mathscr{M}^{(\ell)} | \log | \mathscr{M}^{(\ell)} | + K_\ell \log K_\ell )$. 
    \item \textbf{Signal-level detection:} The optimal assignment of values of $\mathcal{A}_\ell$ to indices in $\hat{\mathcal{I}}_g^{(\ell)}$ can be found via linear search, so the required complexity is $\mathcal{O} (K_\ell ! \cdot K_\ell )$
\end{itemize} Considering that $K_\ell \ll M$ and $| \mathscr{M}^{(\ell)} | \leq M$, $\hat{\mathbf{x}}_g^\text{A-MAP}$ can be obtained after $L$ iterations within a complexity of $\mathcal{O} ( L K M \log M )$. In the last 3rd stage, re-encoding operation to compute $\mathbf{U}_g \hat{\mathbf{x}}_g^\text{A-MAP}$ can be completed with linear complexity, since it is equivalent to the summation of selected columns. Therefore, hypothesis testing can be done with $\mathcal{O} (G)$ complexity. The total complexity of MMSE-A-MAP decoding is $\mathcal{O} ( G L K M \log M )$. In the low-rate regime of interest, the most relevant practical design is single- or two-layer BOSS codes, reducing the complexity to $\mathcal{O}(G K M \log M)$.


\section{Parallel ML Decoder}
In this section, we first show the inapplicability of the original two-stage MAP decoder to the scenario of interest. We, then, propose a two-stage ML decoding architecture which features a parallelization factor of $G$.
\subsection{Limits of the MAP Decoder}
We return to the SIMO uplink problem and consider recovering $\mathbf{x}$ from $\mathbf{y}^{[n]}$ in \eqref{each antenna received vector}. In an analogous fashion to decoding in other two scenarios, we first attempt to remove correlations under each hypothesis:
\begin{equation}
    \mathbf{y}_g^{[n]} = \mathbf{U}_g^\dagger \mathbf{y}^{[n]} = \sqrt{\zeta} h_n \mathbf{x}_g + \Tilde{\mathbf{v}}_n , \label{SIMO decorrelated output}
\end{equation} where $\Tilde{\mathbf{v}}_n = \mathbf{U}_g^\dagger \mathbf{v}_n$. We can formulate the first-stage MAP decoding problem as:
\begin{align*}
    \hat{\mathbf{x}}_g^\text{MAP} &= \underset{\hat{\mathbf{x}}_g \in \mathscr{X}_g}{\arg \max} \, \sum_{n=1}^{N_\text{Rx}} \log \mathbb{P} ( \hat{\mathbf{x}}_g | \mathbf{y}^{[n]} , \mathscr{H}_g ) \\
    &= \underset{\hat{\mathbf{x}}_g \in \mathscr{X}_g}{\arg \max} \, \sum_{n=1}^{N_\text{Rx}} \log \mathbb{P} ( \hat{\mathbf{x}}_g | \mathbf{y}_g^{[n]} ) . \numberthis{\label{SIMO two-stage MAP_first stage}}
\end{align*} The log APP in \eqref{SIMO two-stage MAP_first stage} can be decomposed into the sum of $L$ likelihoods as in \eqref{two-stage MAP_first stage_joint log APP}. At each layer, the MAP decoder identifies $K_\ell$ coefficients that yield the largest likelihood $\mathbb{P} ( y_{g,m}^{[n]} | \hat{x}_{g,m}^{(\ell)} \in \mathcal{A}_\ell )$. For ease of explanation, suppose that each layer's alphabet is a singleton with $\mathcal{A}_\ell = \{ \alpha_\ell \}$. The conditional density functions (pdf's) of elements of $\mathbf{y}_g^{[n]}$ at the $\ell$-th layer's decoding are given by
\begin{equation}
    f_{Y_g^{[n]} | \hat{X}_g^{(\ell)}} ( y_{g,i}^{[n]} | \hat{x}_{g,i}^{(\ell)} ) = \begin{cases}
        \mathcal{CN}(0, \zeta \alpha_\ell^2 + \sigma_v^2), & \text{ if } i \in \mathcal{I}^{(\ell)} \\
        \mathcal{CN}(0, \sigma_v^2), & \text{ else.}
    \end{cases} \label{SIMO conditional pdf}
\end{equation} It can be seen that the element-wise APP $\mathbb{P} ( y_{g,m}^{[n]} | \hat{x}_{g,m}^{(\ell)} \in \mathcal{A}_\ell ) $ cannot be computed without the acquisition of instantaneous channel realizations. The original two-stage MAP decoder becomes inapplicable; instead, we shall consider ML decoding.

\subsection{Divide and Conquer Approach and Two-stage ML Decoder}
Although \eqref{total message vector set} and each of \eqref{sub-message vector set} per $\mathscr{H}_g$ comprise vectors of different dimensions, respectively, in fact, by padding element vectors of each $\mathscr{X}_g$ with zero's, $\mathscr{X}$ can be partitioned as follows
\begin{equation}
    \mathscr{X} = \bigcup_{g=1}^G \Bar{\mathscr{X}}_g,
\end{equation} where
\begin{equation}
    \Bar{\mathscr{X}}_g := \left\{ \begin{bmatrix}
        \mathbf{0}_M^\mathsf{T} & \cdots \mathbf{x}_g^\mathsf{T} & \cdots & \mathbf{0}_M^\mathsf{T} 
    \end{bmatrix} \in \mathbb{C}^N : \mathbf{x}_g \in \mathscr{X}_g \right\} .
\end{equation} Based on this partitioning and independence assumption between receiver antennas, the ML solution can be written as
\begin{align*}
    \hat{\mathbf{x}}^\text{ML} &= \underset{\hat{\mathbf{x}} \in \mathscr{X}}{\arg \max} \, \sum_{n=1}^{N_\text{Rx}} \log \mathbb{P} ( \mathbf{y}^{[n]} | \hat{\mathbf{x}} ) \\
    &= \begin{bmatrix}
        \mathbf{0}_M^\mathsf{T} & \cdots & \hat{\mathbf{x}}_{\hat{g}}^\text{ML} & \cdots & \mathbf{0}_M^\mathsf{T}
    \end{bmatrix}^\mathsf{T} , \numberthis{\label{optimal ML solution}}
\end{align*} where
\begin{equation}
    \hat{\mathbf{x}}_{\hat{g}}^\text{ML} = \underset{\{ \hat{\mathbf{x}}_g^\text{ML} \}_{g \in [G]} }{\arg \max} \, \sum_{n=1}^{N_\text{Rx}} \log \mathbb{P} ( \mathbf{y}^{[n]}  | \hat{\mathbf{x}}_g^\text{ML} ) \label{sub-ML solution-block identification}
\end{equation} and
\begin{equation}
    \hat{\mathbf{x}}_g^\text{ML} = \underset{\hat{\mathbf{x}}_g \in \mathscr{X}_g}{\arg \max} \sum_{n=1}^{N_\text{Rx}} \log \mathbb{P} ( \mathbf{y}^{[n]} | \hat{\mathbf{x}}_g , \mathbf{A} \mathbf{x} = \mathbf{U}_g \mathbf{x}_g ) . \label{sub-ML solution-under hypothesis}
\end{equation} That is, $\hat{\mathbf{x}}_g^\text{ML}$ is the ML estimate under $\mathscr{H}_g$. Then, the optimal ML solution can still be found by solving \eqref{sub-ML solution-under hypothesis} and \eqref{sub-ML solution-block identification} separately in each stage. The first stage returns a set of $G$ length-$M$ vectors, $\{ \hat{\mathbf{x}}_1^\text{ML} , \hat{\mathbf{x}}_2^\text{ML}, \dots , \hat{\mathbf{x}}_G^\text{ML} \}$. It is apparent that each $\hat{\mathbf{x}}_g^\text{ML}$ can be obtained without any information exchange with computations of $\{ \hat{\mathbf{x}}_j^\text{ML} \}_{j \in [G] \backslash \{g \} }$ under other hypothesis. Therefore, the first stage can be efficiently implemented with a parallelization factor of the block size $G$, considerably reducing the decoding latency. The second-stage decoding is to identify the true block index utilizing the first-stage outputs. By padding $\hat{\mathbf{x}}_{\hat{g}}^text{ML}$ with $G - 1$ all zero vectors, the decoder returns $\hat{\mathbf{x}}^\text{ML}$ in \eqref{optimal ML solution}.

\section{Approximate ML Decoding}
In the previous section, we have demonstrated achieving the ML solution by resolving multiple ML problems over a reduced search space for each hypothesis, thus granting a high degree of parallelization. However, addressing each reduced problem still remains problematic. In this section, we present efficient methodologies for quasi-ML decoding. Our approach hinges on presuming Gaussianity in the received signals across individual antennas post-decorrelation. Subsequently, leveraging the insights gleaned from the study of structural properties of BOSS codes, we introduce the second major contribution of this paper: a sphere-decoding-based quasi-ML solution tailored for practical deployment scenarios.

\subsection{Quasi-ML Decoding}
In the absence of CSI, our decoding algorithm relies on energy-detection. Consequently, decoding multi-layer BOSS codes with alphabets of varying power levels poses significant challenges and falls outside the practical scope. We, hence, confine our discussion to a single-layer BOSS code with a singleton alphabet: $L = 1$, $K = K_1$, and $\mathcal{A} = \mathcal{A}_1 = \{ \alpha \}$. The first-stage decoding problem in \eqref{two-stage ML-first stage} can be re-written as 
\begin{equation}
    \hat{\mathbf{x}}_g^\text{ML} = \underset{\hat{\mathbf{x}}_g \in \mathscr{X}_g}{\arg \max} \, \sum_{n=1}^{N_\text{Rx}} \log \mathbb{P} ( \mathbf{y}_g^{[n]} | \hat{\mathbf{x}}_g ) ,
\end{equation} where $\mathbf{y}_g^{[n]}$ is the decorrelated $n$-th antenna output defined in \eqref{SIMO decorrelated output}. Invoking the conditional pdf's of elements of $\mathbf{y}_g^{[n]}$, which are given in \eqref{SIMO conditional pdf}, it can be seen that $\mathbf{y}_g^{[n]}$ conforms to a degenerate multivariate Gaussian distribution. This is because its non-zero elements at indices of $\mathcal{I}$ are perturbed by the same channel realization, thereby introducing correlation among them. Unfortunately, the joint density of $\mathbf{y}_g^{[n]}$ cannot be computed. Following a line of reasoning similar to that employed in the development of MMSE-A-MAP algorithm, and taking into account the sparsity of $\mathbf{x}$ and the low transmission power required by BOSS codes, we overlook correlations among the few non-zero components. As such, we assume that elements of $\mathbf{y}_g^{[n]}$ are independent complex Gaussian with variances $\text{Var} \{ y_{g, i}^{[n]} \} \in \{ \zeta \alpha^2 + \sigma_v^2 , \sigma_v^2 \}$. We further postulate that $\mathbb{E} \{ \mathbf{y}_g^{[n]} \mathbf{y}_g^{[n]}\ctran \, \}$ is the covariance matrix of $\mathbf{y}_g^{[n]}$ despite non-zero off-diagonal elements. Under these assumptions, the conditional joint density of $\mathbf{y}_g^{[n]}$ can be expressed in a closed form, and we have the quasi-ML decoding solution as follows:
\begin{align*}
    & \hat{\mathbf{x}}_g^\text{q-ML} \\
    &= \underset{\hat{\mathbf{x}}_g \in \mathscr{X}_g}{\arg \max} \, \sum_{n=1}^{N_\text{Rx}} \log \Biggl[ \frac{1}{\pi^M \det \left( \mathbb{E} \left\{ \mathbf{y}_g^{[n]} {\mathbf{y}_g^{[n]}}\ctran  \right\} \right) } \\
    & \quad \quad \quad \cdot \exp \left( - {\mathbf{y}_g^{[n]}}\ctran \, \mathbb{E} \left\{ \mathbf{y}_g^{[n]} {\mathbf{y}_g^{[n]}}\ctran \right\}^{-1} \mathbf{y}_g^{[n]} \right) \Biggr] \\
    & = \underset{\hat{\mathbf{x}}_g \in \mathscr{X}_g}{\arg \max} \, \sum_{n=1}^{N_\text{Rx}} \log \Biggl[ \frac{1}{\pi^M \det ( \zeta \hat{\mathbf{x}}_g \hat{\mathbf{x}}_g^\mathsf{T} + \sigma_v^2 \mathbf{I}_M)} \\
    & \quad \quad \quad \cdot \exp \left( - {\mathbf{y}_g^{[n]}}\ctran \left( \zeta \hat{\mathbf{x}}_g \hat{\mathbf{x}}_g^\mathsf{T} + \sigma_v^2 \mathbf{I}_M \right)^{-1} \mathbf{y}_g^{[n]} \right) \Biggr] \\
    &\overset{(a)}{=} \underset{\hat{\mathbf{x}}_g \in \mathscr{X}_g}{\arg \max} \, \sum_{n=1}^{N_\text{Rx}} \log \Biggl[ \frac{1}{\pi^M ( \sigma_v^2 + K \zeta \alpha^2 )} \\
    & \quad \quad \quad \cdot \exp \left( - {\mathbf{y}_g^{[n]}}\ctran \left( \zeta \hat{\mathbf{x}}_g \hat{\mathbf{x}}_g^\mathsf{T} + \sigma_v^2 \mathbf{I}_M \right)^{-1} \mathbf{y}_g^{[n]} \right) \Biggr] \\
    &\overset{(b)}{=} \underset{\hat{\mathbf{x}}_g \in \mathscr{X}_g}{\arg \max} \, \sum_{n=1}^{N_\text{Rx}} \log \Biggl[ \frac{1}{\pi^M ( \sigma_v^2 + K \zeta \alpha^2 )} \\
    & \quad \quad \quad  \cdot \exp \Biggl( - {\mathbf{y}_g^{[n]}}\ctran \Biggl( \frac{1}{\sigma_v^2} \mathbf{I}_M - \frac{\frac{\zeta}{\sigma_v^2} \hat{\mathbf{x}}_g \hat{\mathbf{x}}_g^\mathsf{T}}{\sigma_v^2 + K \zeta \alpha^2} \Biggr) \mathbf{y}_g^{[n]} \Biggr) \Biggr] , \numberthis{\label{quasi-ML first-stage v1}}
\end{align*} where the labeled equalities follow from: $(a)$ the Sylvester determinant identity, i.e., $\det ( \zeta \hat{\mathbf{x}}_g \hat{\mathbf{x}}_g^\mathsf{T} + \sigma_v^2 \mathbf{I}_M ) = \det ( \sigma_v^2 + \zeta \hat{\mathbf{x}}_g^\mathsf{T} \hat{\mathbf{x}}_g ) $ and $(b)$ the Sherman-Morrison formula, i.e., $\left( \mathbf{I}_M + \frac{\zeta}{\sigma_v^2} \hat{\mathbf{x}}_g \hat{\mathbf{x}}_g^\mathsf{T} \right)^{-1} = \mathbf{I}_M - \frac{\frac{\zeta}{\sigma_v^2} \mathbf{I}_M ( \hat{\mathbf{x}}_g \hat{\mathbf{x}}_g^\mathsf{T} ) \mathbf{I}_M}{1 + \frac{\zeta}{\sigma_v^2} \hat{\mathbf{x}}_g^\mathsf{T} \mathbf{I}_M \hat{\mathbf{x}}_g } $. By disregarding the constants and terms irrelevant of $\hat{\mathbf{x}}_g$, \eqref{quasi-ML first-stage v1} reduces to the following:
\begin{align*}
    \hat{\mathbf{x}}_g^\text{q-ML} &= \underset{\hat{\mathbf{x}}_g \in \mathscr{X}_g}{\arg \max} \, \sum_{n=1}^{N_\text{Rx}} {\mathbf{y}_g^{[n]}}\ctran \, 
    \hat{\mathbf{x}}_g \hat{\mathbf{x}}_g^\mathsf{T} \mathbf{y}_g^{[n]} \\
    &= \underset{\hat{\mathbf{x}}_g \in \mathscr{X}_g}{\arg \max} \, \hat{\mathbf{x}}_g^\mathsf{T} \left( \sum_{n=1}^{N_\text{Rx}} \mathbf{y}_g^{[n]} {\mathbf{y}_g^{[n]}}\ctran \right) \hat{\mathbf{x}}_g , \numberthis{\label{quasi-ML first-stage v2}}
\end{align*} where the equality arises from the invariance of the trace under cyclic permutations. By stacking $N_\text{Rx}$ decorrelated antenna outputs, let us define an $M \times N_\text{Rx}$ matrix $\mathbf{Y}_g$ as $\mathbf{Y}_g = \begin{bmatrix}
    \mathbf{y}_g^{[1]} & \mathbf{y}_g^{[2]} & \cdots & \mathbf{y}_g^{[N_{\text{Rx}}]} 
\end{bmatrix}$. Then, \eqref{quasi-ML first-stage v2} can be re-written as:
\begin{equation}
    \hat{\mathbf{x}}_g^\text{q-ML} = \underset{\hat{\mathbf{x}}_g \in \mathscr{X}_g}{\arg \max} \, \hat{\mathbf{x}}_g^\mathsf{T} \left( \mathbf{Y}_g \mathbf{Y}_g^\dagger \right) \hat{\mathbf{x}}_g = \underset{\hat{\mathbf{x}}_g \in \mathscr{X}_g}{\arg \max} \, \| \mathbf{Y}_g^\dagger \hat{\mathbf{x}}_g \|_2^2 . \label{quasi-ML first-stage v3}
\end{equation} The final equivalent problem in \eqref{quasi-ML first-stage v3} is intriguing, for it tells that the quasi-ML estimate can be obtained by computing and comparing an inner product of $\mathbf{Y}_g^\dagger \hat{\mathbf{x}}_g$ with itself. This result is intuitively understanding, as the decoder searches for a vector that is most-correlated with the received signals.

\subsection{Two-stage Non-coherent Sphere Decoding} 
\begin{algorithm}
    \caption{\textbf{Algorithm II: } Non-Coherent Sphere Decoding}
    \begin{algorithmic}[1]
        \Require $N_\text{Rx}$ antenna outputs $\left\{ \mathbf{y}^{[1]} , \mathbf{y}^{[2]}, \dots , \mathbf{y}^{[N_\text{Rx}]} \right\}$ and sphere set size parameter $T \geq K$
        \Ensure sparse message vector $\hat{\mathbf{x}}^\text{NSD} \in \mathcal{A}^N$ with $\| \hat{\mathbf{x}}^\text{NSD} \|_0 = K$
        \For{$g = 1, \dots, G$}
            \State Decorrelate each antenna's received signal and obtain $\{ \mathbf{y}_g^{[n]} \}_{n \in N_\text{Rx}}$
            \State Compute $\Tilde{\mathbf{K}}_g = \frac{1}{N_\text{Rx}} \sum_{n=1}^{N_\text{Rx}} \mathbf{y}_g^{[n]} {\mathbf{y}_g^{[n]}}^\dagger$ 
            \State Calculate $\mathscr{R}_{g, i}$ and $\mathscr{C}_{g, i}$ as the sum of the $K$ largest elements of each row and column, respectively
            \State Find $T$ largest $\mathcal{Y}_{g, i} = \mathscr{R}_{g, i} + \mathscr{C}_{g, i}$ and their indices
            \State Construct a sphere set $\mathcal{S}_{g, T}$ of $\hat{\mathbf{x}}_g$'s whose supports are a subset of the previously found $T$ indices
            \State Perform quasi-ML decoding over $\mathcal{S}_{g, T}$ and obtain $\hat{\mathbf{x}}_g^\text{NSD}$
        \EndFor
        \State Perform hypothesis testing and identify the most-likely candidate $\hat{\mathbf{x}}_{\hat{g}}^\text{NSD}$
        \State \Return $\hat{\mathbf{x}}^\text{NSD} = \begin{bmatrix}
            \mathbf{0}_M^\mathsf{T} & \cdots & {\hat{\mathbf{x}}_{\hat{g}}^\text{NSD}}\tran & \cdots & \mathbf{0}_M^\mathsf{T}
        \end{bmatrix}$
    \end{algorithmic}
\end{algorithm}
The quasi-ML problem \eqref{quasi-ML first-stage v2} can be rewritten as
\begin{equation}
    \hat{\mathbf{x}}_g^\text{q-ML} = \underset{\hat{\mathbf{x}}_g \in \mathscr{X}_g}{\arg \max} \, \hat{\mathbf{x}}_g^\mathsf{T} \Tilde{\mathbf{K}}_g \hat{\mathbf{x}}_g = \underset{\substack{ {\hat{\mathcal{I}}_g = \text{supp}(\hat{\mathbf{x}}_g )}\\{\hat{\mathbf{x}}_g \in \mathscr{X}_g}  }}{\arg \max} \, \sum_{i, j \in \hat{\mathcal{I}}_g} ( \Tilde{\mathbf{K}}_g)_{i, j} , \label{quasi-ML first-stage v4}
\end{equation} where $\Tilde{\mathbf{K}}_g = \sum_{n=1}^{N_\text{Rx}} \mathbf{y}_g^{[n]} { \mathbf{y}_g^{[n]} }^\dagger$. That is, solving for $\hat{\mathbf{x}}_g^\text{q-ML}$ is equivalent to finding $K^2$ largest elements of $\Tilde{\mathbf{K}}_g$.

Let us denote by $\Dot{\mathbf{x}}_g$ and $\Dot{\mathcal{I}}_g$ the true sub-message vector and its support under $\mathscr{H}_g$, respectively. Each $n$-th outer product constituting $\Tilde{\mathbf{K}}_g$ can be decomposed as:
\begin{multline}
    \mathbf{y}_g^{[n]} {\mathbf{y}_g^{[n]}}^\dagger \\
    \!\!\!\!=\! \zeta \| h_n \|_2^2 \Dot{\mathbf{x}}_g \Dot{\mathbf{x}}_g^\mathsf{T} \!+\! \sqrt{\zeta} h_n \Dot{\mathbf{x}}_g \Tilde{\mathbf{v}}_n^\dagger \!+\! \sqrt{\zeta} h_n^* \Tilde{\mathbf{v}}_n \Dot{\mathbf{x}}_g^\mathsf{T} \!+\! \Tilde{\mathbf{v}}_n \Tilde{\mathbf{v}}_n^\dagger. \label{outer product decomposition}
\end{multline} The second term features non-zero rows, while the third term contains non-zero columns, both indexed by $\Dot{\mathcal{I}}_g$ and representing scaled replicas of $\Tilde{\mathbf{v}}_n^\dagger$ and $\Tilde{\mathbf{v}}_n$, respectively. Therefore, even if the noise-only component, $\Tilde{\mathbf{v}}_n \Tilde{\mathbf{v}}_n^\dagger$, is neglected, the second and third term may contribute to non-zero elements.


While the presence of undesired contributions from noise adds complexity to the task of identifying entries with large magnitudes, sparsity proves advantageous in addressing this challenge. We propose an non-coherent sphere decoding (NSD) algorithm for BOSS codes. Upon computing $\Tilde{\mathbf{K}}_g$, we calculate the sum of $K$ largest elements of each row and column, denoted by $\mathscr{R}_{g, i}$ and $\mathscr{C}_{g, i}$, respectively: 
\begin{equation}
    \mathscr{R}_{g, i} := \sum_{j=1}^K ( \Tilde{\mathbf{K}}_g )_{i, \pi_{i}' (j)} \text{ and } \mathscr{C}_{g, i} := \sum_{j=1}^K ( \Tilde{\mathbf{K}}_g )_{\pi_{i}''(j), i} , 
\end{equation} where $\pi_{i}' (j)$ and $\pi_{i}''(j)$ represent the sorted indices of the elements of the $i$-th row and column of $\Tilde{\mathbf{K}}_g$, respectively. The decoder, then, identifies $T \geq K$ largest $\mathcal{Y}_{g, i} := \mathscr{R}_{g, i} + \mathscr{C}_{g, i}$, and their indices: $\left\{ \mu(1), \mu(2), \dots , \mu(T) \right\}$. That is, if $i < j$, then $\mathcal{Y}_{g, \mu(i)} \geq \mathcal{Y}_{g, \mu(j)} $ for all $i, j \in [T]$. The decoder constructs a sphere set, $\mathcal{S}_{g, T}$, of distinct message vector estimates whose supports are a subset of $\{ \mu(1), \mu(2), \dots, \mu(T) \}$:
\begin{equation}
    \mathcal{S}_{g, T} := \{ \hat{\mathbf{x}}_g \in \mathscr{X}_g : \hat{\mathcal{I}}_g \subseteq \{ \mu(1) , \mu(2) , \dots , \mu(T) \} \}.
\end{equation} Finally, the decoder performs quasi-ML decoding in \eqref{quasi-ML first-stage v4} over the above sphere set:
\begin{equation}
    \hat{\mathbf{x}}_g^\text{NSD} = \underset{\hat{\mathbf{x}}_g \in \mathcal{S}_{g, T}}{\arg \max} \, \hat{\mathbf{x}}_g^\mathsf{T} \Tilde{\mathbf{K}}_g \hat{\mathbf{x}}_g .
\end{equation} It is evident that the number of vectors to consider reduces from $2^{\lfloor \log_2 \left( \binom{M}{K} \right) \rfloor }$ to $\binom{T}{K}$, and the computational complexity decreases accordingly.

The decoder identifies the most-probable sub-message vector estimate via hypothesis testing:
\begin{equation}
    \hat{\mathbf{x}}_{\hat{g}}^\text{NSD} = \underset{\hat{\mathbf{x}}_g^\text{NSD}: g \in [G]}{\arg \max} \, {\hat{\mathbf{x}}_g^\text{NSD}}\tran \, \Tilde{\mathbf{K}}_g \hat{\mathbf{x}}_g^\text{NSD} .
\end{equation} The overall NSD algorithm is presented in Algorithm 2. 

\section{Simulation Results}
This section presents block error rate (BLER) simulation results to validate the error correction capabilities of BOSS codes under two proposed decoding algorithms.
\subsection{MMSE-A-MAP Decoding}
We model the channel using a seven-tap model with an exponential power delay profile, simulating an indoor environment. Each channel coefficient is represented as:
\begin{equation}
    \xi_i = \frac{1}{Z} e^{i - 1}, \quad i = 1, 2, \dots, 7, 
\end{equation} where $Z$ is a normalizing constant to ensure $\| \bm{\xi} \|_2^2 = 1$. The fast Fourier transform size is set to $64$, with $48$ subcarriers reserved for data. Therefore, each codeword is carried by $\lceil \frac{M}{48} \rceil$ OFDM blocks. Given the primary focus on transmitting short packets, we assume that the channel coherence period is always greater than the code blocklength. 

\begin{figure}[h]
    \centering
    \includegraphics[width=0.8\columnwidth]{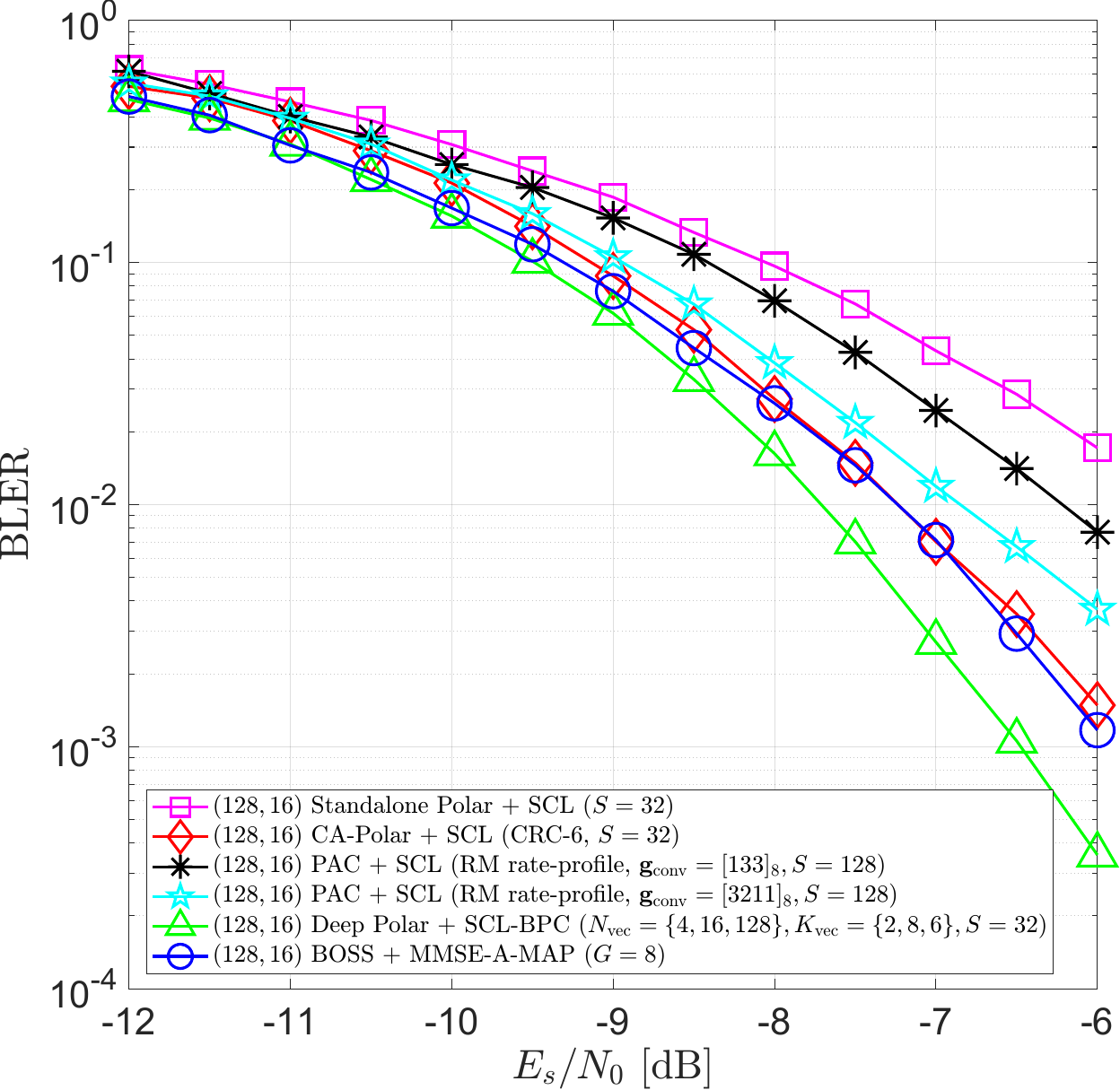}
    \caption{BLER performance comparison of different $(128, 16)$ codes in a seven-tap Rayleigh channel.}
    \label{fig:BLER_Plot1}
    \vspace{-2.5mm}
\end{figure}
We evaluated various $(M = 128, B = 16)$ channel codes. Their design parameters and respective decoding algorithms are provided below.
\begin{itemize}
    \item \textbf{BOSS codes:} We simulated a two-layer BOSS code with $G = 8$; symmetric sparsity $K_1 = K_2 = 1$; singleton alphabets $\mathcal{A}_1 = \{1\}$ and $\mathcal{A}_2 = \{ -1 \}$; and candidate set cardinalities $| \mathscr{M}^{(1)} | = 128$ and $| \mathscr{M}^{(2)} | = 64$.
    \item \textbf{(CA-) Polar codes:} We considered a 5G NR polar code \cite{3GPP_TS_38_212} with the binary phase-shift keying (BPSK) modulation. An optional CRC outer code was employed with the CRC-6 polynomial. Both standalone and CA-polar codes were evaluated under successive cancellation list (SCL) decoding with a list size of $S = 32$ \cite{Tal15_Polar_SCL}.
    \item \textbf{PAC codes:} We adopted the Reed-Muller rate-profile \cite{Ariakn19_PAC} and employed two different generator polynomials: $\mathbf{g}_\text{conv} = 133$ and $\mathbf{g}_\text{conv} = 3211$ in octal notation. Both codes were BPSK-modulated and decoded using the SCL-based algorithm with a list size of $S = 128$ \cite{Yao23_PAC_SCL}. 
    \item \textbf{Deep polar codes:} A deep polar code is a novel variant of pre-transformed polar codes where the pre-transform consists of multi-layered nested polar encoding \cite{Choi24_Deep_Polar}. We tested a three-layer deep polar code with $N_\text{vec} = \{4, 8, 128\}$ and $K_\text{vec} = \{2, 8, 6 \}$ under BPSK signaling . For decoding, we used the SCL with backpropagation parity-check decoder (SCL-BPC) \cite{Choi24_Deep_Polar} with $S = 32$.
\end{itemize} For SCL-based decoding of polar code variants, channel log-likelihood ratios (LLRs) were computed by factoring in the exact values of frequency-domain channel coefficients, assuming perfect CSI. The $i$-th bit-channel LLR is given by
\begin{equation}
    \mathsf{L}_i \!:=\! \log \frac{\mathbb{P} (y_i | c_i = +1 )}{\mathbb{P} (y_i | c_i = -1 )} \!=\! \frac{4 (\Re \{ y_i \} \!\cdot\! \Re \{ \lambda_i \} \!+\! \Im \{ y_i \} \!\cdot\! \Im \{ \lambda_i \} )  }{\sigma_v^2} .
\end{equation} The list size $S$ was chosen to achieve saturating ML-like performance for polar code variants.

The performance comparison results in Fig. \ref{fig:BLER_Plot1} indicate that BOSS codes under MMSE-A-MAP achieve performance comparable to CA-polar codes, while deep polar codes exhibit approximately a $0.5$ dB coding gain at a BLER of $10^{-3}$. 

\subsection{Concatenation of CRC Outer Code and List Decoding}
\begin{figure*}[!t]
    \centering
    \subfloat[]{\includegraphics[width=0.25\textwidth]{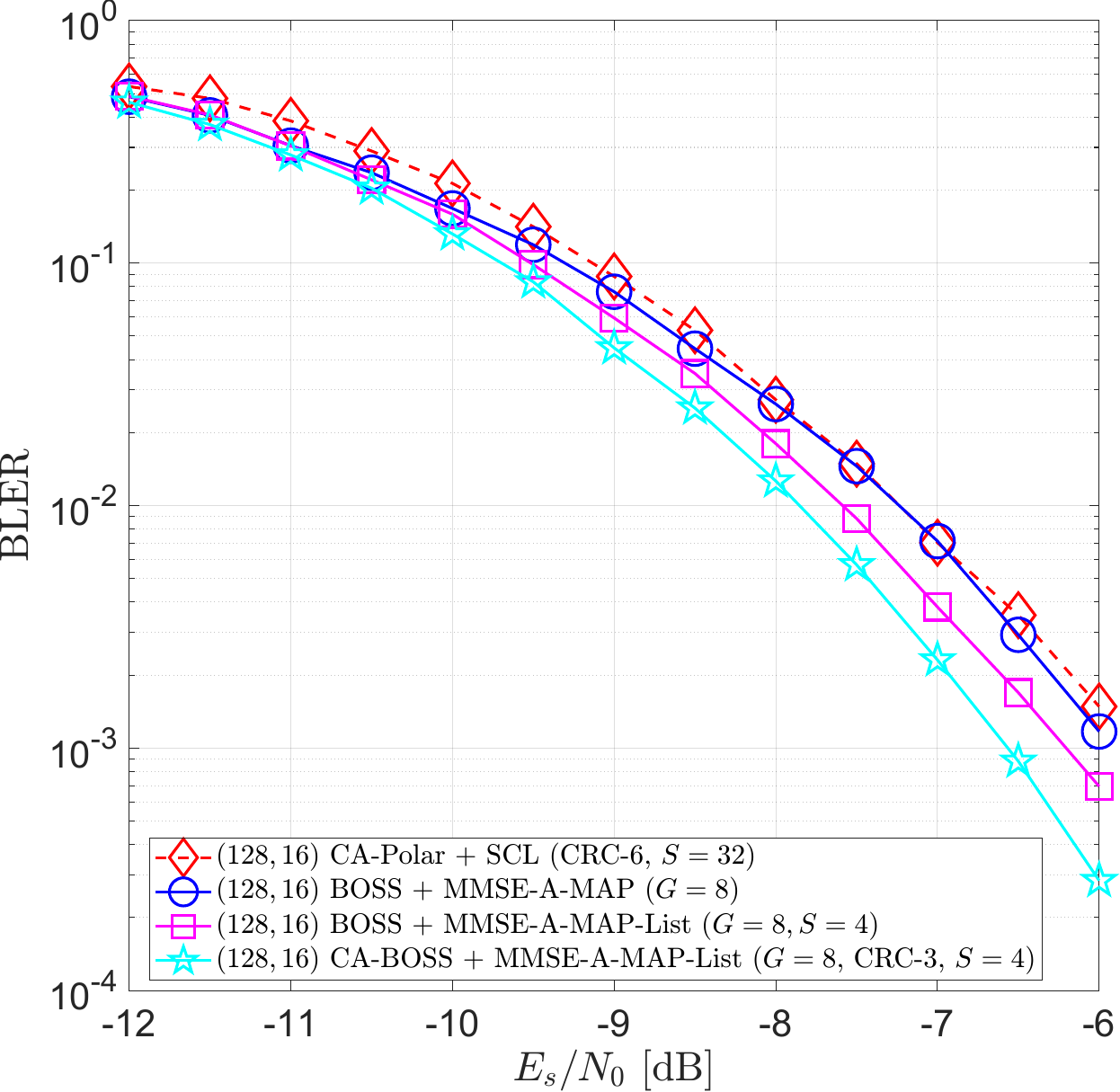}%
    \label{fig2_a}}
    \hfil
    \subfloat[]{\includegraphics[width=0.25\textwidth]{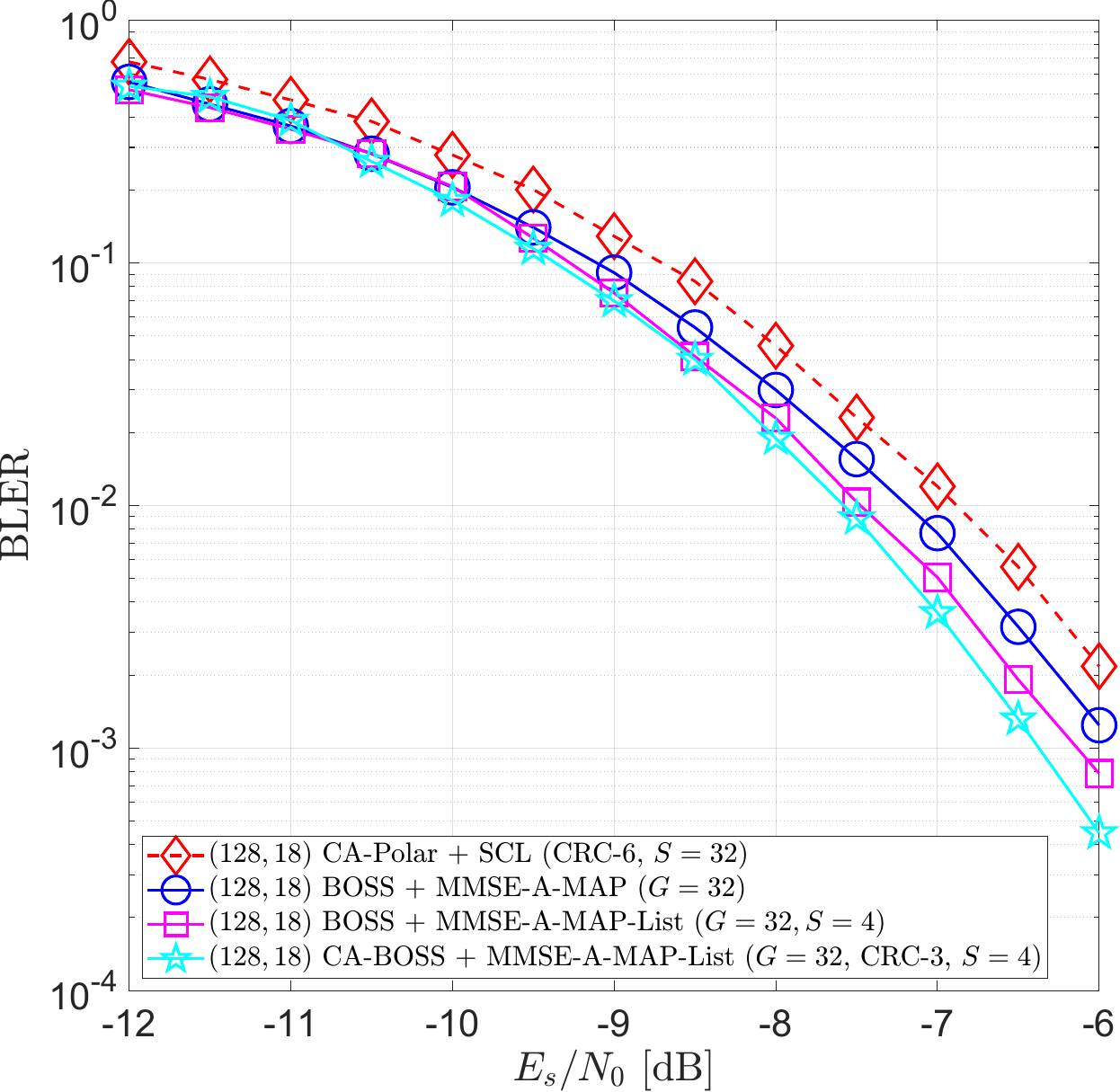}%
    \label{fig2_b}}
    \hfil
    \subfloat[]{\includegraphics[width=0.25\textwidth]{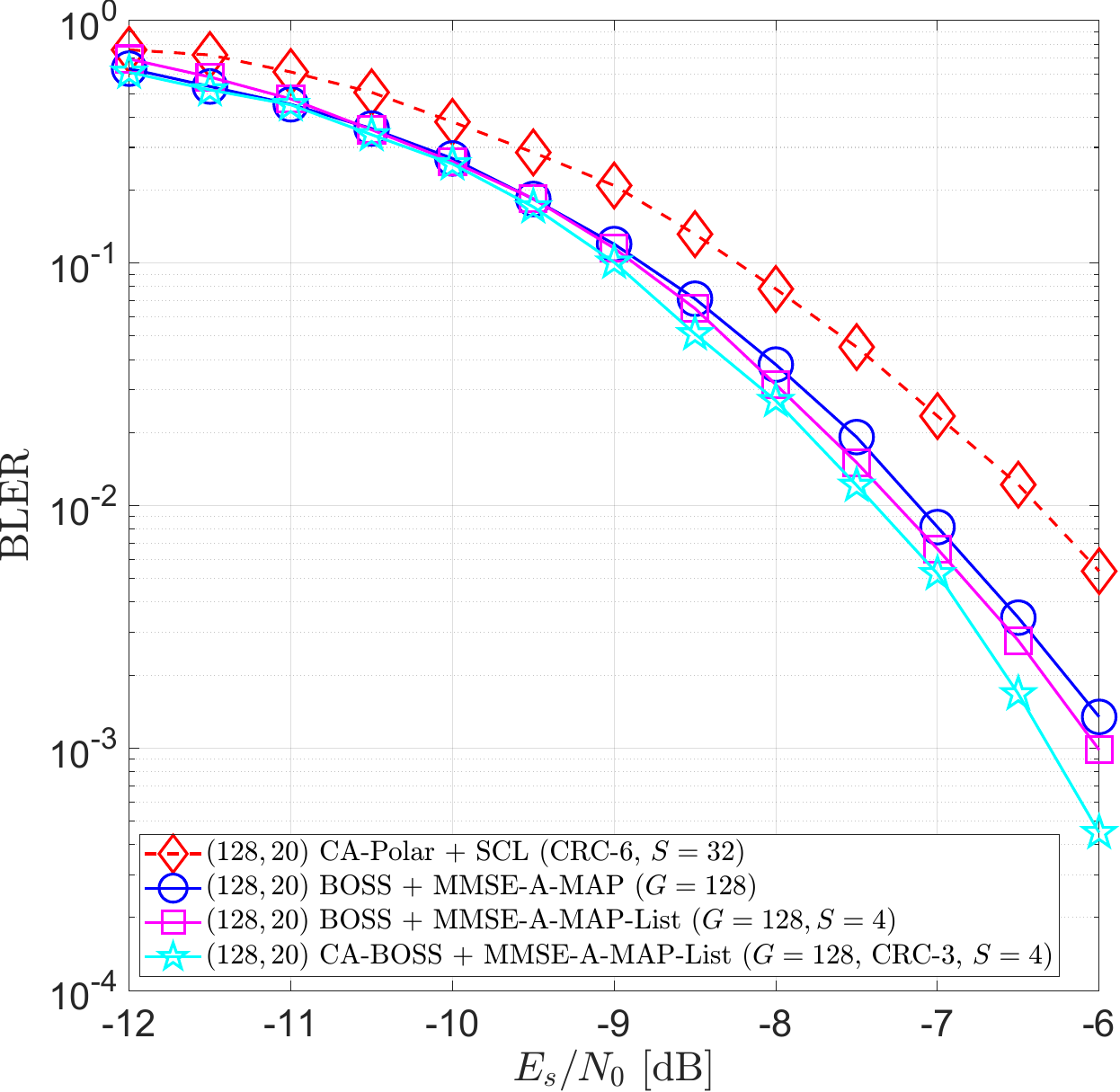}%
    \label{fig2_c}}

    \subfloat[]{\includegraphics[width=0.25\textwidth]{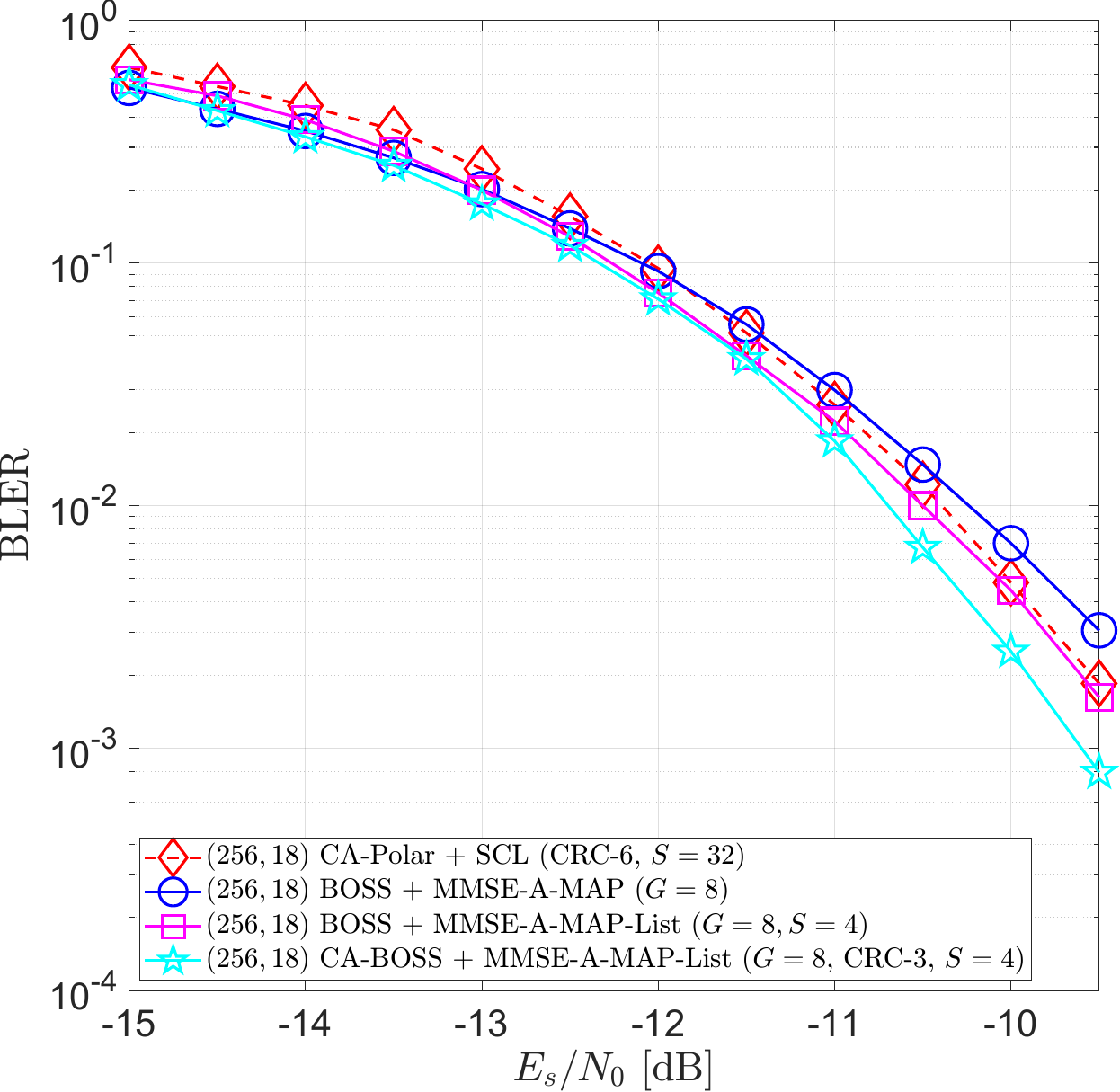}%
    \label{fig2_d}}
    \hfil
    \subfloat[]{\includegraphics[width=0.25\textwidth]{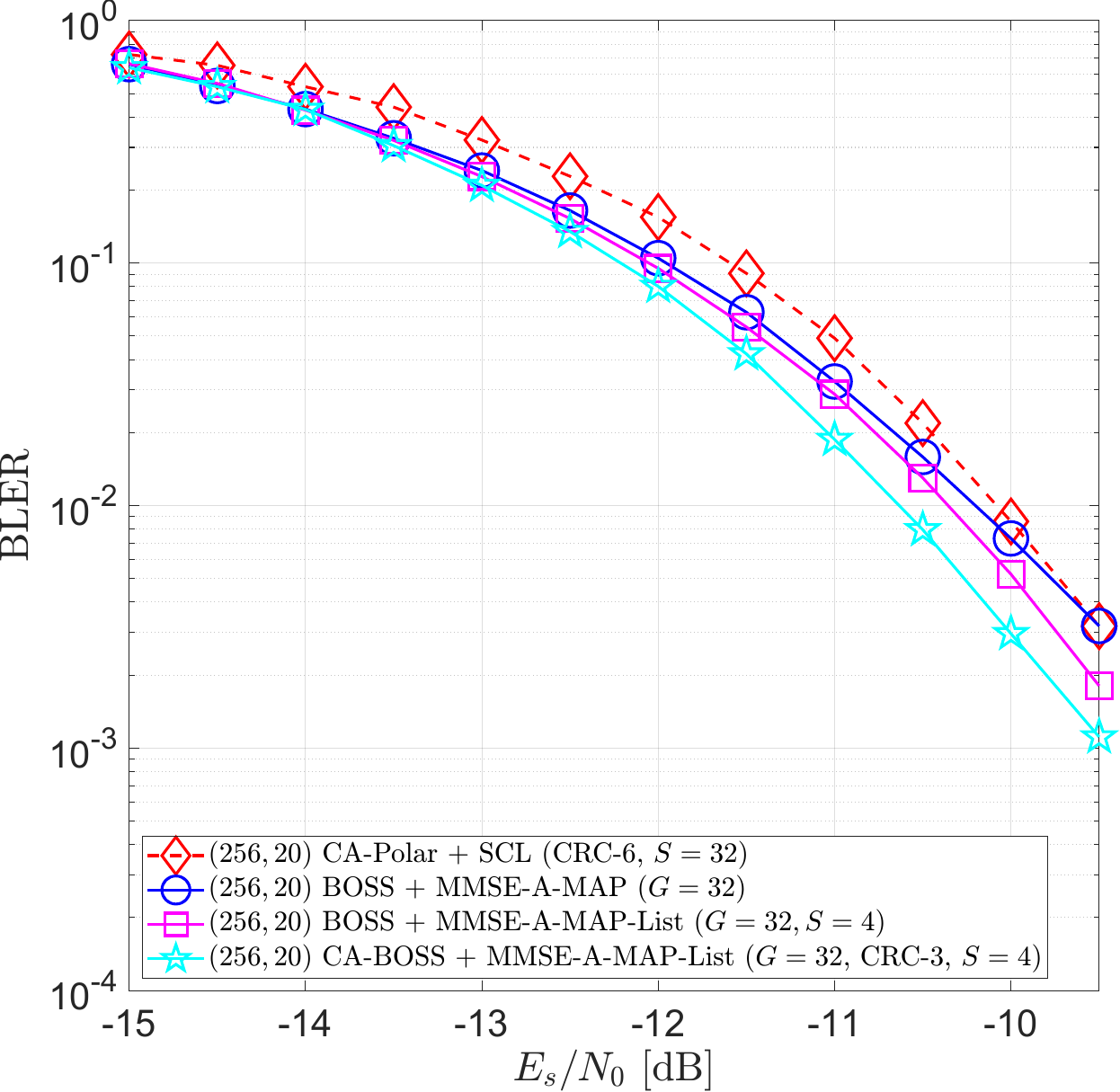}%
    \label{fig2_e}}
    \hfil
    \subfloat[]{\includegraphics[width=0.25\textwidth]{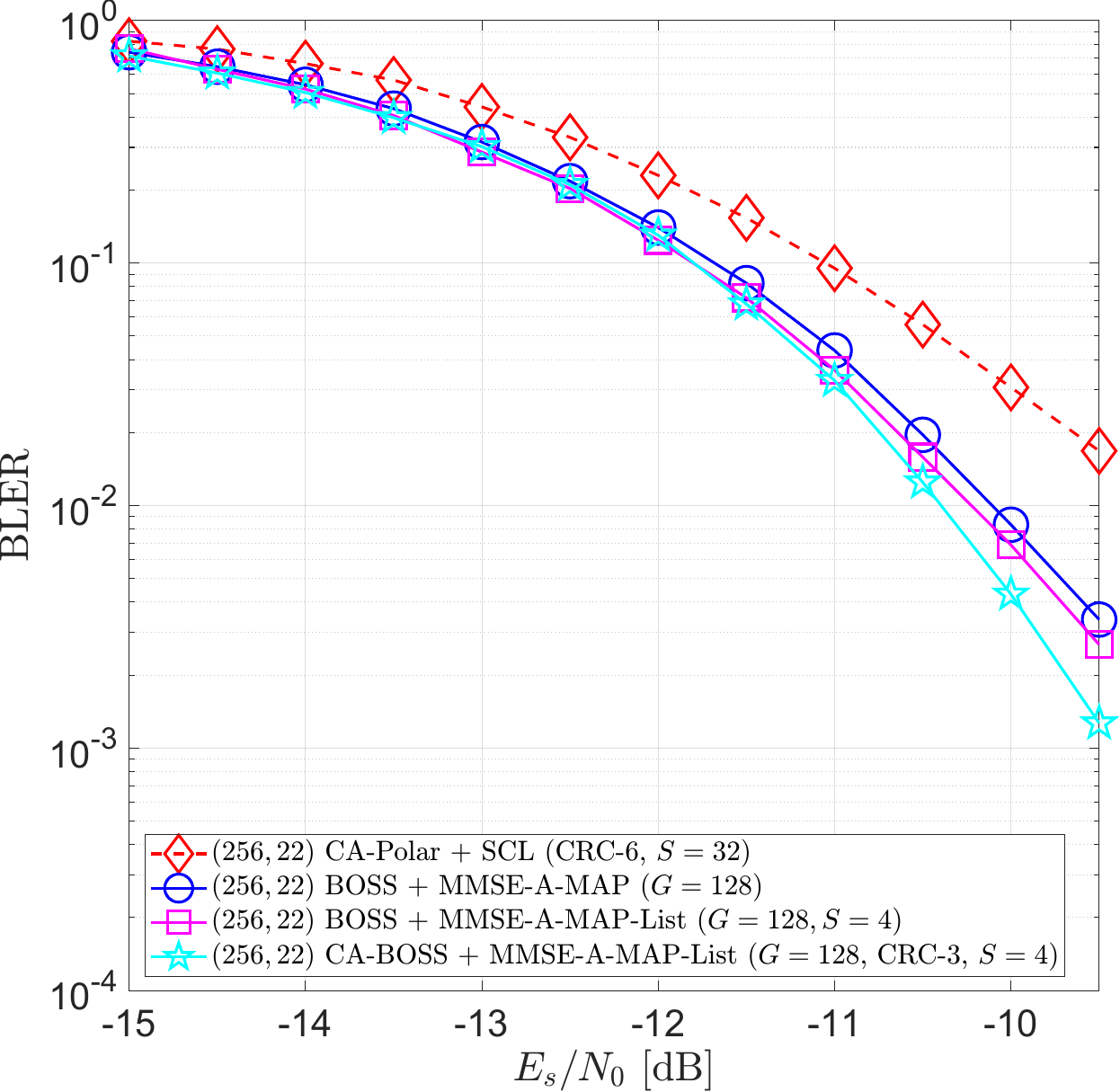}%
    \label{fig2_f}}
    \caption{Enhanced performance of CA-BOSS codes with $M \in \{128, 256\}$ and varying block number $G \in \{8, 32, 128\}$.}
    \label{fig:BLER_Plot2}
\end{figure*}

\cite{BOSS_AWGN_journal} demonstrated that the error-correction capability of BOSS codes can be further improved through concatenation with a CRC outer code, resulting in CA-BOSS codes, and by incorporating a list decoder into the original decoder framework. We adopt these techniques for the MMSE-A-MAP decoder: in the second stage, the decoder generates a \textit{list} of CRC-valid combinations of non-zero support and block index.

Fig. \ref{fig:BLER_Plot2} illustrates the performance of BOSS and CA-BOSS codes with lengths $M \in \{128, 256 \}$ and varying block sizes $G \in \{8, 32, 128\}$ under MMSE-A-MAP-List decoding with a list size of $S = 4$, compared to CA-polar codes of respective coding rates. It is observed that CA-BOSS codes with CRC-3 outperform CA-polar codes in all cases by approximately $0.4 \sim 1.0$ dB, depending on the configurations.

\subsection{Quasi-ML versus NSD}
\begin{figure*}[!t]
    \centering
    \subfloat[]{\includegraphics[width=0.23\textwidth]{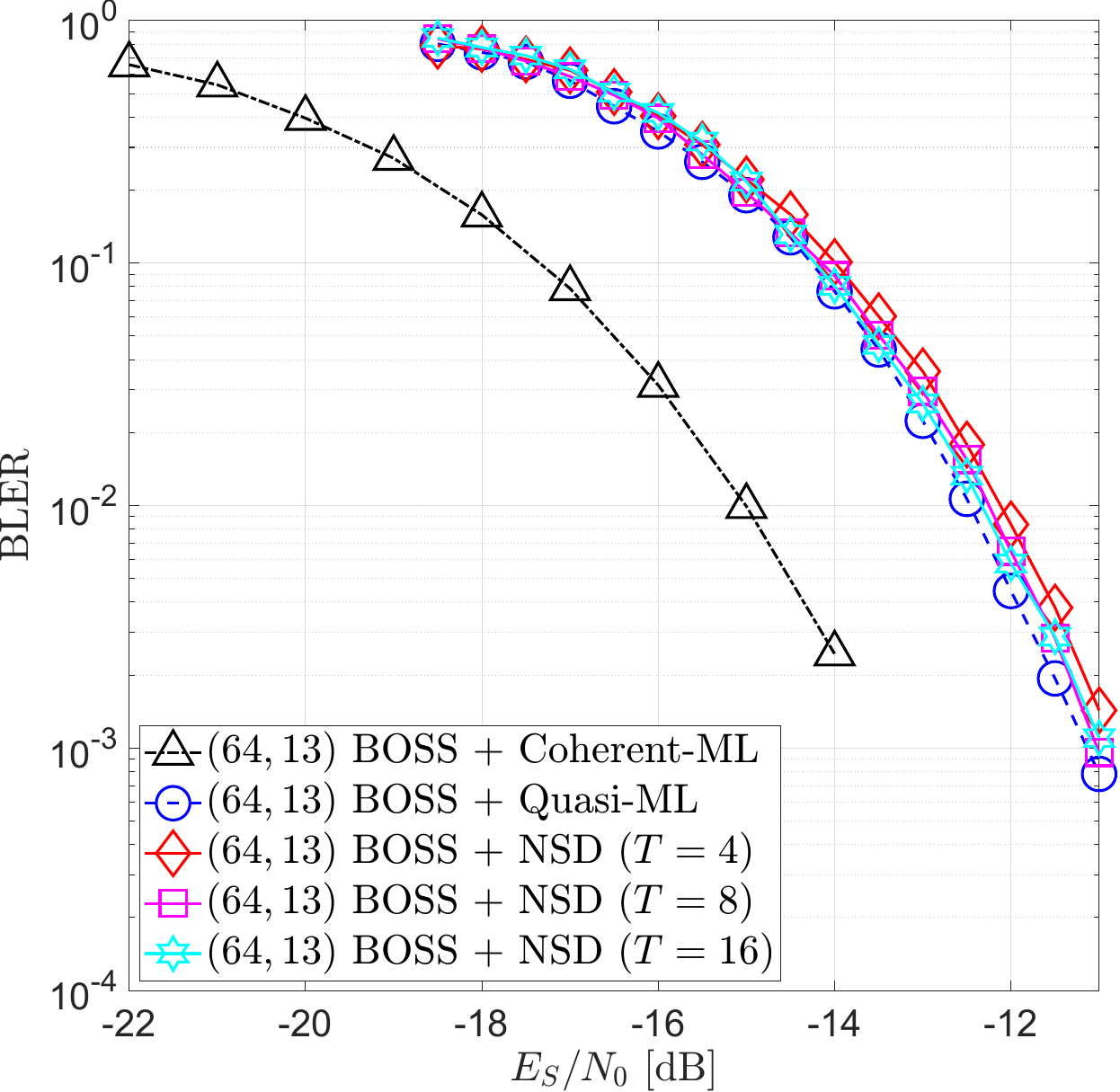}%
    \label{subfig:BLER_Plot3_N_Rx16}}
    \hfil
    \subfloat[]{\includegraphics[width=0.23\textwidth]{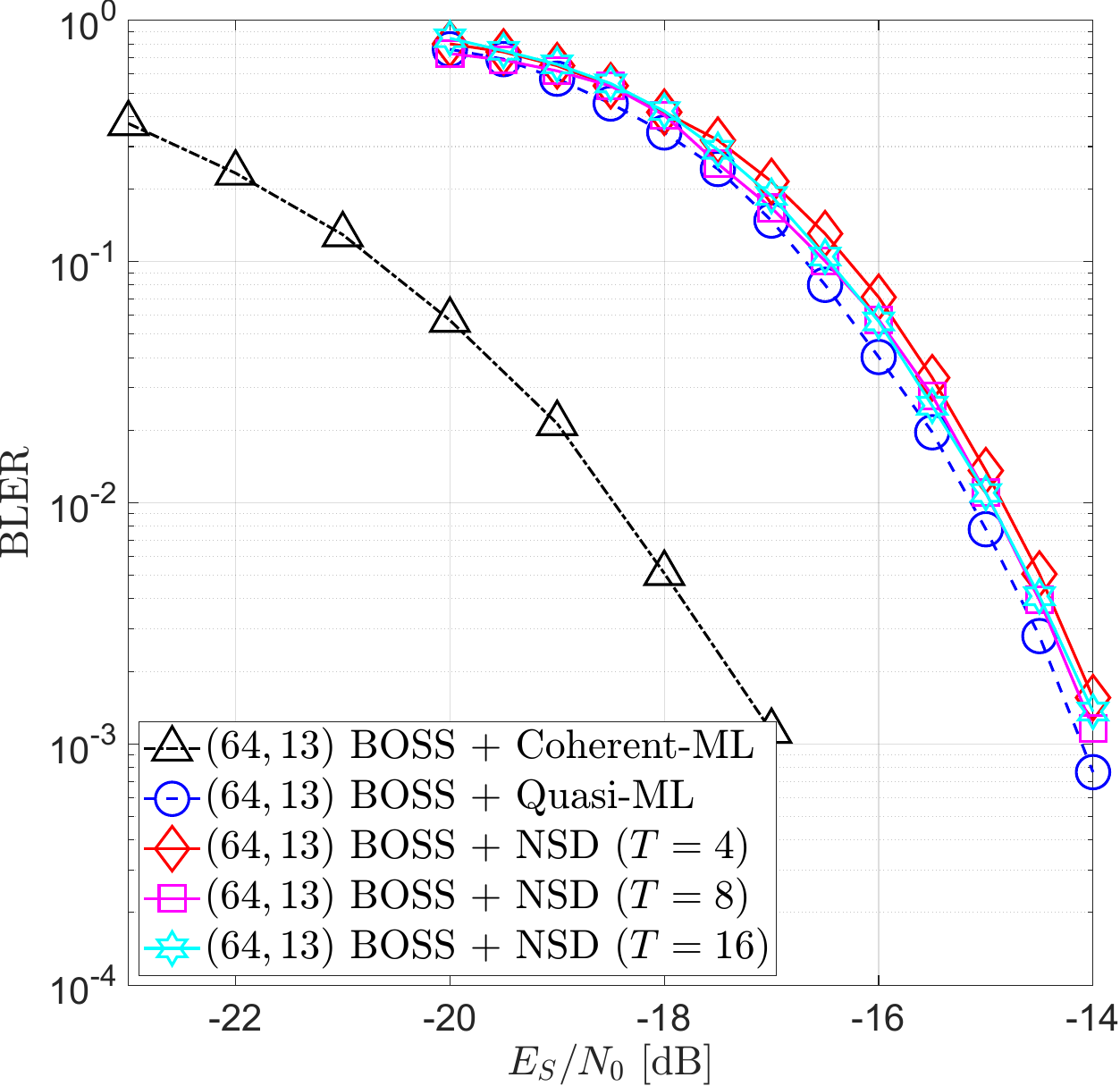}%
    \label{subfig:BLER_Plot3_N_Rx32}}
    \hfil
    \subfloat[]{\includegraphics[width=0.23\textwidth]{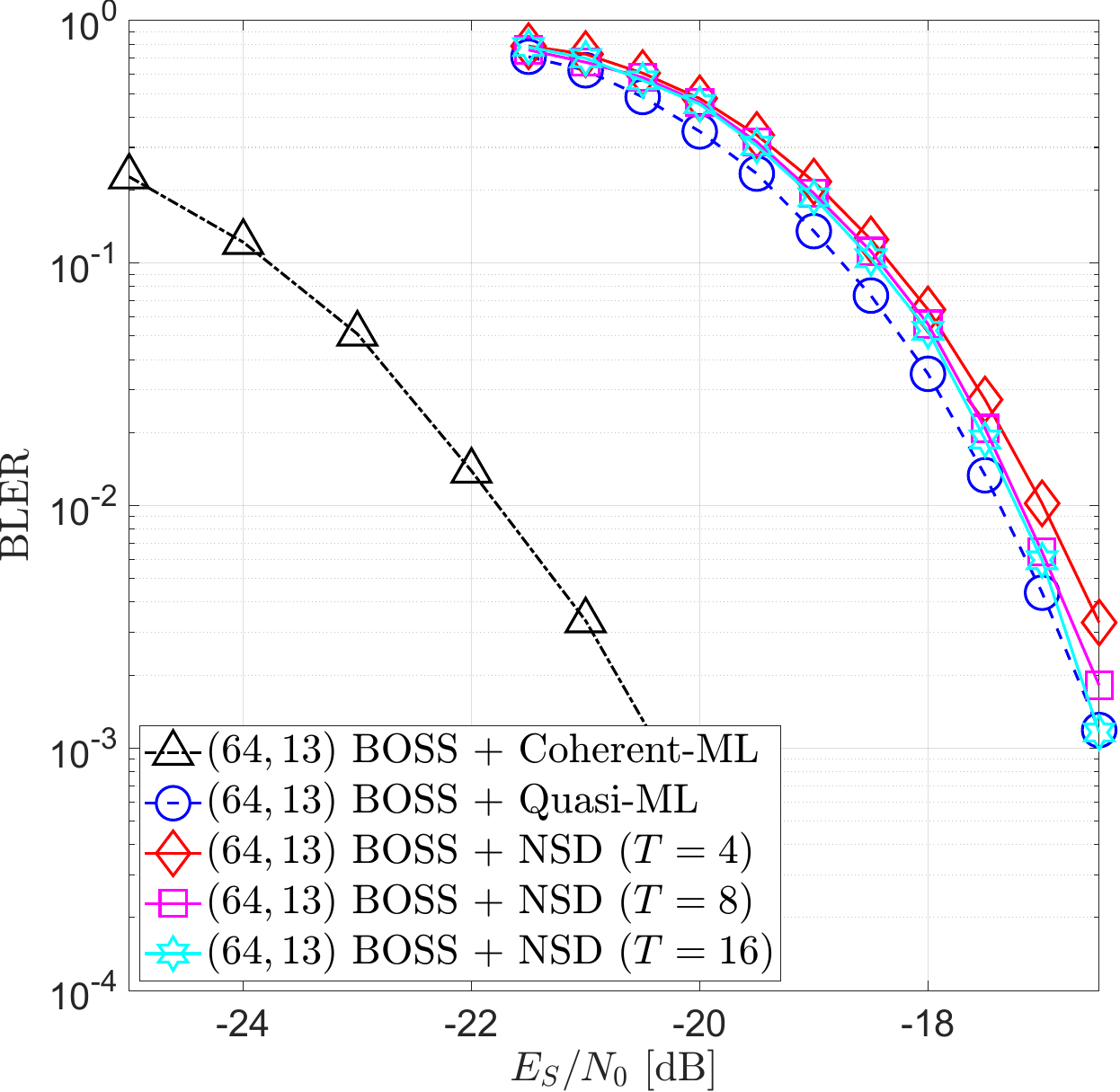}%
    \label{subfig:BLER_Plot3_N_Rx64}}
    \caption{Performance of $(M = 64, G = 8, K = 2)$ single-layer BOSS codes under different decoders when the number of receiver antennas are: (a) $N_\text{Rx} = 16$, (b) $N_\text{Rx} = 32$, and (c) $N_\text{Rx} = 64$.}
    \label{fig:BLER_Plot3}
\end{figure*}

We first considered a single-layer length-$64$ BOSS code with $K = 2$. With $G = 8$, each codeword conveys $B = \log_2 (8) + \lfloor \log_2 (\binom{64}{2} ) \rfloor = 13$ bits. We compared the proposed NSD algorithm with different sphere set parameters $T \in \{4, 8, 16 \}$ to the quasi-ML decoder in terms of BLER. Fig. \ref{fig:BLER_Plot3} provides simulation results for settings with an increasing number of receiver antennas: $N_\text{Rx} \in \{16, 32, 64\}$. It can be seen that even with a small value of $T = 8$, the NSD decoder performs very closely to the quasi-ML decoder. For the purpose of benchmark, Fig. \ref{fig:BLER_Plot3} also plots the BLER results of ML decoding with perfect CSI, i.e., coherent ML decoding.

\begin{figure}[!t]
    \centering
    \includegraphics[width=0.7\columnwidth]{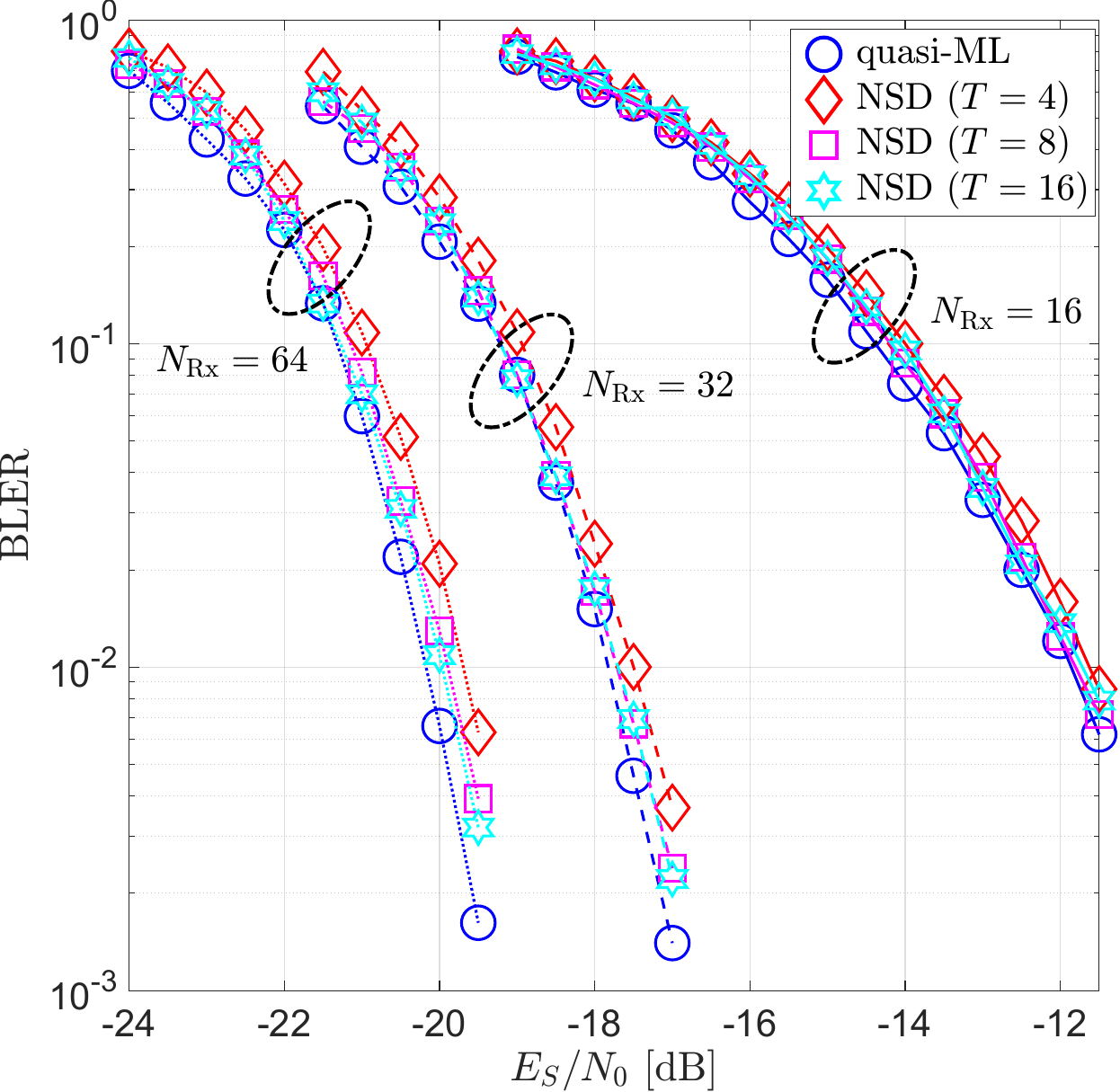}
    \caption{$(M = 128, G = 8, K = 2)$ single-layer BOSS codes under quasi-ML decoding and NSD.}
    \label{fig:BLER_Plot4}
\end{figure}

Fig. \ref{fig:BLER_Plot4} evaluates $(M = 128, G = 8, K = 2)$ single-layer BOSS codes under quasi-ML decoding and NSD with $T \in \{4, 8, 16 \}$. Each codeword carries $B = \log_2 (8) + \lfloor \log_2 ( \binom{128}{2} ) \rfloor = 15$ bits. Across three cases with $N_\text{Rx} \in \{16, 32, 64 \}$, NSD achieves comparable performance to quasi-ML decoding. Notably, in the given configuration, NSD with $T = 8$ considers $\binom{8}{2} = 28$ codewords per hypothesis, a significant reduction from $2^{\lfloor \log_2 ( \binom{128}{2} ) \rfloor } = 4096$ of the quasi-ML counterpart, representing approximately a $99 \%$ reduction.

\begin{figure}[h]
    \centering
    \includegraphics[width = 0.65\columnwidth]{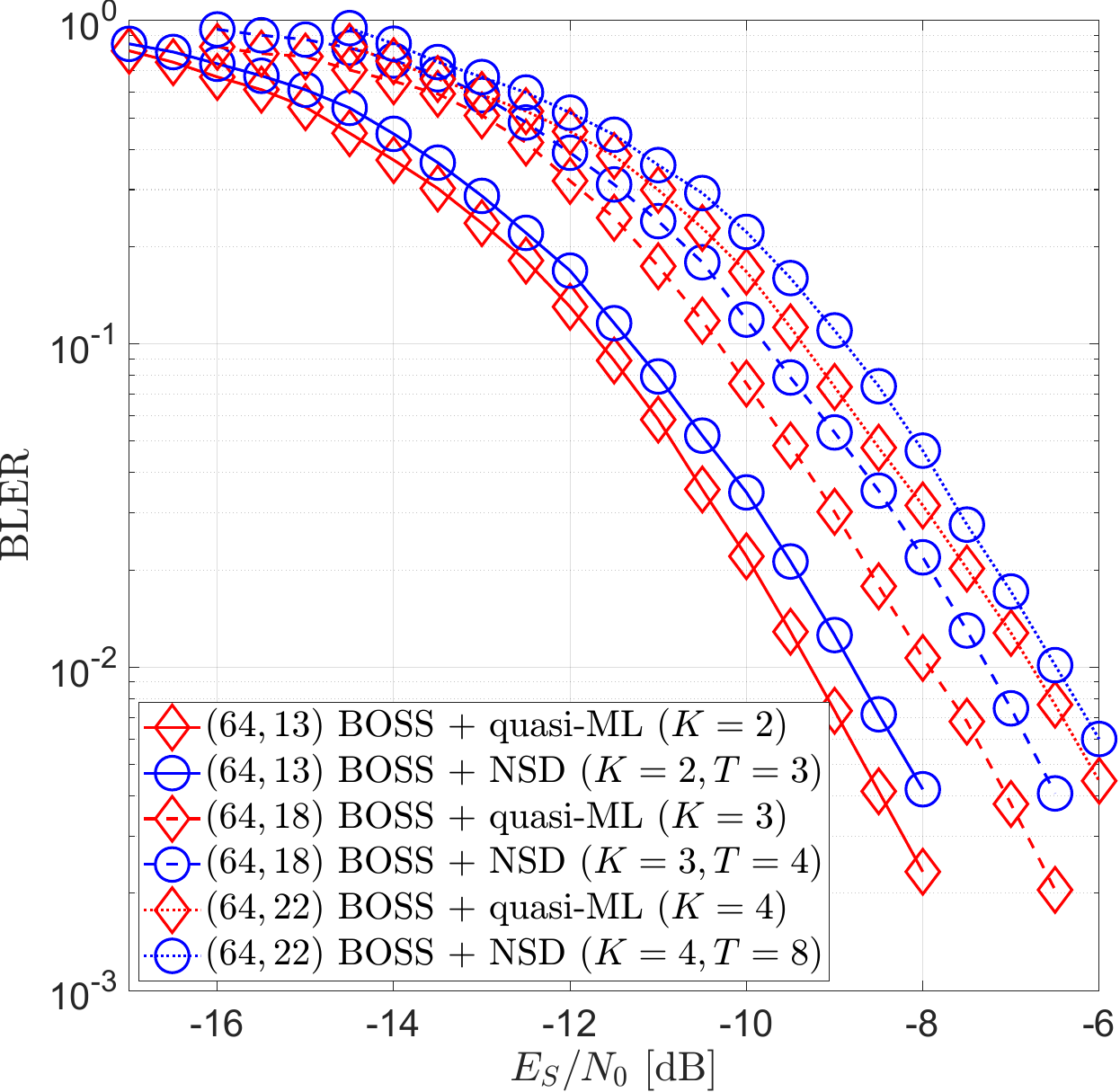}
    \caption{$(M = 64, G = 8)$ single-layer BOSS codes with varying sparsity $K \in \{2, 3, 4\}$ when $N_\text{Rx} = 8$.}
    \label{fig:BLER_Plot5}
\end{figure}

Fig. \ref{fig:BLER_Plot5} assesses $(M = 64, G = 8)$ single-layer BOSS codes while varying the number of non-zero entries, $K \in \{2, 3, 4\}$, under the set-up $N_\text{Rx} = 8$. Results confirm the effectiveness of the proposed NSD algorithm: even with $T$ slightly larger than $K$, NSD performs within $0.5$ dB of quasi-ML decoding.

\section{SDR Testbed}
\begin{figure}[!t]
    \centering
    \includegraphics[width=0.65\columnwidth]{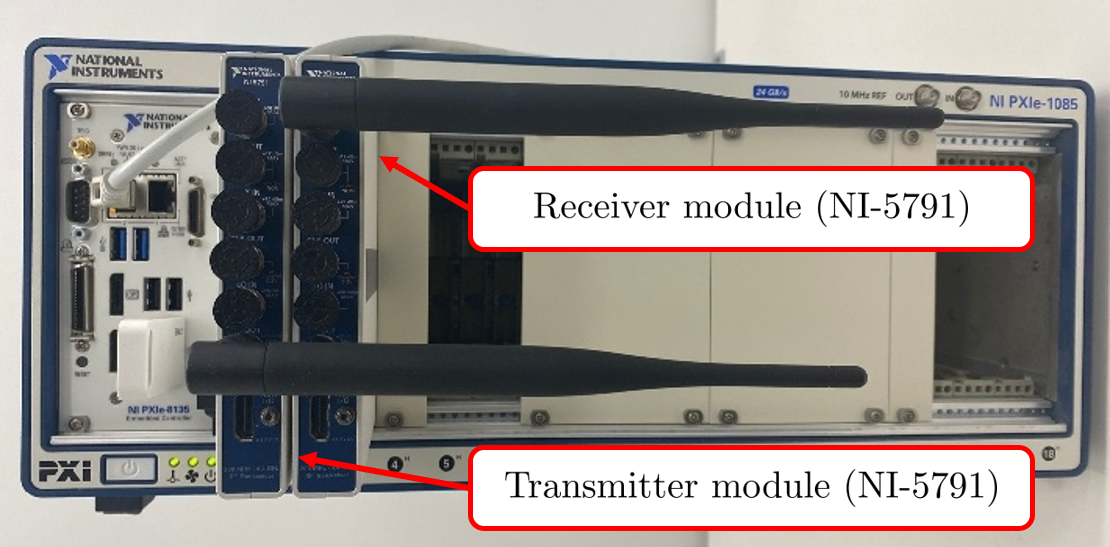}
    \caption{SDR-based SISO testbed with the NI-PXIe platform.}
    \label{fig:testbed}
    \vspace{-3mm}
\end{figure}



\subsection{Testbed Set-up}
Our SDR testbed shown in Fig. \ref{fig:testbed} is constructed on the NI-PXIe modular platform, where the NI-PXIe-1085 chassis accommodates two single-antenna NI-5791 modules. One module serves as the transmitter, while the other functions as the receiver. Software-based digital signal processing is employed in the transmitter (DSP-Tx) and receiver (DSP-RX) sides. In DSP-Tx, random binary data are encoded into BOSS symbols, followed by OFDM modulation alongside preambles. Subsequently, after upsampling and pulse-shaping, the symbols undergo power amplification before transmission via RF signals. In DSP-Rx, the received RF signals are subjected to matched-filtering and downsampling, after which synchronization and carrier-frequency offset take place utilizing preambles. Finally, demodulated OFDM symbols are decoded using the MMSE-A-MAP algorithm, returning the estimation of the original binary data. NI LabVIEW is used to streamline the processes of DSP-Tx and DSP-Rx. Details of the simulation configuration are provided in Table I.
\begin{table}[]
    \centering
    \caption{SDR simulation set-up with BOSS code parameters}
    \begin{tabular}{l c}
        \toprule 
        \multicolumn{2}{c}{\textbf{Simulation Configuration}} \\
        \midrule
             Training sequence                                      &       Schmidl-Cox \cite{Cox97_OFDM_preamble} \\
             Radio frequency                                        &       2.4 GHz \\
             Bandwidth                                              &       20 MHz \\
             Transmit power                                         &       -10 dBm \\
             FFT size                                               &       64 \\
             Cyclic prefix length                                   &       16 \\
             Number of null sub-carriers                            &       4 \\
             Upsampling rate                                        &       20 \\
             Pulse-shaping filter                                   &       Root-raised-cosine \\
             Roll-off factor                                        &       0.5 \\
         \bottomrule
    \end{tabular}
    \vspace{-3mm}
    \label{tab:my_label}
\end{table}
\subsection{Experimental Results}
\begin{figure}[!h]
    \centering
    \subfloat[]{\includegraphics[width=0.4\columnwidth]{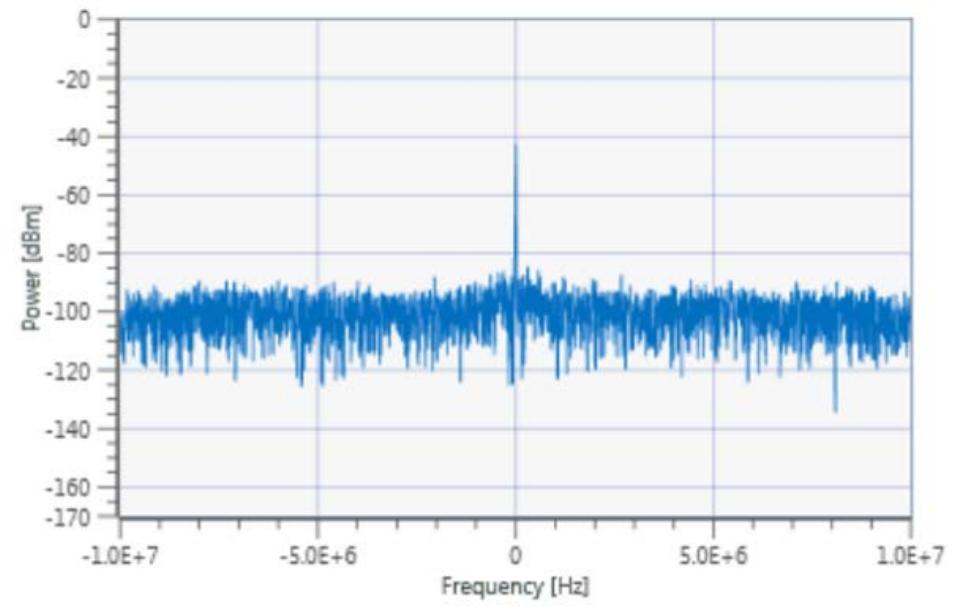}\label{psd_transmitted}}
    \hfill
    \subfloat[]{\includegraphics[width=0.4\columnwidth]{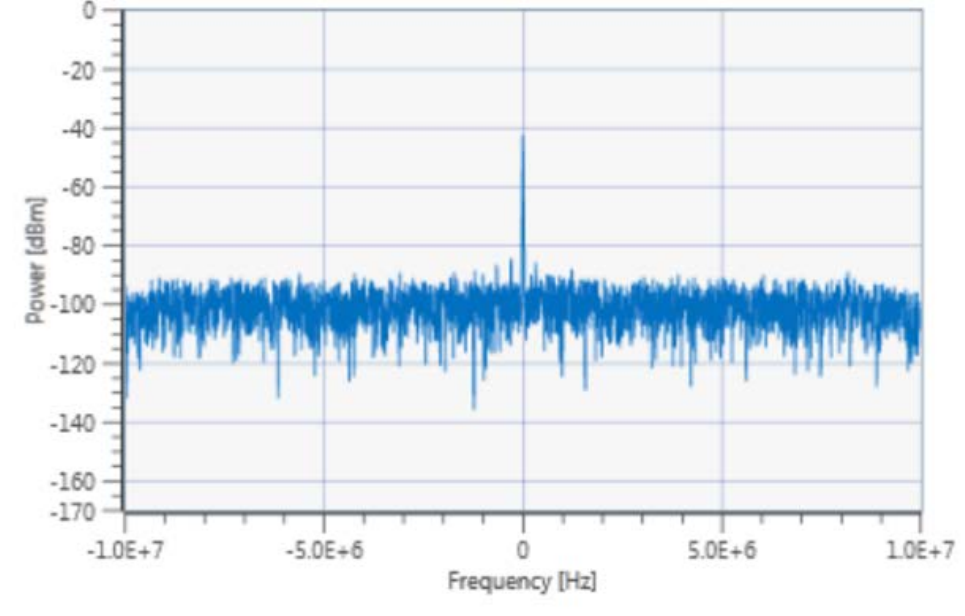}\label{psd_noise_only}}
    \caption{Snapshot of the received signal PSD when a BOSS codeword is transmitted (a), and noise-only conditions (b).}
    \label{fig:power_spectral_density}
    \vspace{-2.5mm}
\end{figure}
We evaluated $(M = 129, G = 4)$ two-layer BOSS code with $K_1 = K_2 = 1$; and $\mathcal{A}_1 = \{ 1 \}$; $\mathcal{A}_2 = \{ -1 \}$; and $| \mathscr{M}^{(1)} | = 129$ and $| \mathscr{M}^{(2)} | = 128$. Hence, a codeword carries $\lfloor \log_2 (129) \rfloor + \log_2(128) = 14$ bits. In real-time SISO simulations with the specified configuration, 221 errors were detected out of 100,000 codewords transmitted. This corresponds to a BLER of 0.002. Additionally, the power spectral density (PSD) of the received signal, as shown in Fig. \ref{psd_transmitted}, closely mirrors the spectrum shown in Fig. \ref{psd_noise_only}, which represents the noise-only PSD in the absence of transmitted signals. This means that even when the transmitted signals are nearly submerged in the noise, our BOSS decoder is still able to recover the original data. This promising result not only validates the feasibility of the MMSE-A-MAP algorithm but also suggests the potential for low-power communications utilizing BOSS codes. 

\section{Conclusion}
In this paper, we have extended the applicability of BOSS codes to practical fading environments and proposed two powerful, easily-parallelizable decoding algorithms. Our investigation reveals that the MMSE-A-MAP decoder demonstrates remarkable robustness against noise and fading perturbations, delivering superior performance compared to CA-polar codes in the SISO-OFDM system. Additionally, the NSD algorithm for the uplink SIMO system capitalizes on the structural features of the BOSS code, offering comparable performance to quasi-ML decoding while significantly reducing complexity. Finally, we have implemented the MMSE-A-MAP algorithm on the SDR platform for feasibility verification, Empirical results confirm the potential of BOSS codes as a promising candidate for future HRLLC and Ambient IoT applications.


While polar codes exhibit excellent performance for blocklengths of less than a thousand, scaling to longer length poses challenges. As blocklength increases, the baseline SCL decoder requires larger list sizes to achieve satisfactory ML-like performance. This combination of of increased lengths and augmented list sizes results in heightened complexity. Addressing this challenge, a concatenated scheme of BOSS and non-binary low-density parity-check (NB-LDPC) codes has been proposed \cite{Lee24_BOSS_LDPC}. BOSS-LDPC leverages complementary strengths of both codes, compensating for the susceptibility of LDPC codes to short cycles in short-to-moderate blocklengths and the performance degradation of BOSS codes with multiple layers. This synergistic combination delivers superior performance over both CA-polar and standalone LDPC codes. Therefore, extending the study of BOSS-LDPC codes to fading environments presents a promising avenue for future research.

\bibliographystyle{IEEEtran}
\bibliography{main}


\vfill

\end{document}